\documentclass[review,3p,12pt]{elsarticle}

\usepackage{amssymb}
\usepackage{amsmath}
\usepackage[ruled,linesnumbered]{algorithm2e}
\usepackage{setspace}
\usepackage{siunitx}
\usepackage{lineno}

\usepackage{multirow}
\usepackage{booktabs}
\usepackage{overpic}
\usepackage[colorlinks,linkcolor=blue,anchorcolor=blue,citecolor=blue]{hyperref}
\usepackage{etoolbox}
\usepackage{sectsty}
\usepackage[dvipsnames]{xcolor}
\usepackage[final]{changes}
\makeatletter
\ifbool{Changes@optiondraft}{\newcommand{\delcr}{\\}\newcommand{\delamp}{&}\newcommand{\delsp}{\ }}{\newcommand{\delcr}{}\newcommand{\delamp}{}\newcommand{\delsp}{}\renewcommand{\added}[2][]{#2}\renewcommand{\deleted}[2][]{\@bsphack\@esphack\ifhmode\ifnum\lastnodetype=-1 \expandafter\expandafter\expandafter\ignorespaces\fi\fi}}
\makeatother

\def\onedot{$\mathsurround0pt\ldotp$}
\def\cdddot#1{% three dots 
  \mathbin{\vcenter{\baselineskip.67ex
    \hbox{\onedot}\hbox{\onedot}\hbox{\onedot}%
  }}%
}

\journal{Elsevier}

\begin{document}

\begin{frontmatter}


  \title{Towards end-to-end optimization in multimaterial 3D printing}

  \pdfstringdefDisableCommands{%
    \def\corref#1{}%
  }

  \author[cornell]{Xue-Ling Luo}
  % \ead{xl2254@cornell.edu}
  \author[cornell]{Steven Yang}
  % \ead{sjy32@cornell.edu}
  \author[usc]{Jingye Tan}
  % \ead{jingyeta@usc.edu}
  \author[cornell]{Robert F. Shepherd}
  % \ead{rfs247@cornell.edu}
  \author[technion,cornell]{Noy Cohen}
  % \ead{noyco@technion.ac.il}
  \author[cornell,pasteur]{Nikolaos Bouklas\corref{cor1}}
  \ead{nb589@cornell.edu}

  \cortext[cor1]{Corresponding author.}
  
  \affiliation[cornell]{addressline={Sibley School of Mechanical and Aerospace Engineering, Cornell University},
              city={Ithaca},
              % postcode={14850},
              state={NY},
              country={USA}}

   \affiliation[usc]{addressline={Department of Aerospace \& Mechanical Engineering, University of Southern California},
              city={Los Angeles},
              % postcode={90007},
              state={CA},
              country={USA}}
              
   \affiliation[technion]{addressline={Department of Materials Science and Engineering, Technion - Israel Institute of Technology},
              city={Haifa},
              % postcode={3200003},
              country={Israel}}

    \affiliation[pasteur]{addressline={Pasteur Labs},
              city={Brooklyn},
              state={NY},
              country={USA}}

  \begin{abstract}
    Multimaterial 3D printing enables the fabrication of functionally graded components, but optimizing their spatial material distribution alongside structural topology remains a formidable challenge due to high-dimensional design spaces and complex constitutive modeling. \deleted{This paper presents an end-to-end computational framework integrating sparsified physics-augmented neural networks with finite-element-based topology optimization.} \added{Using neural network constitutive laws in topology optimization requires their tangents and design sensitivities, which current approaches obtain by automatic differentiation through the network or derivation by hand. This paper combines sparsification, which reduces a trained physics-augmented neural network to a compact, interpretable closed-form law, with symbolic differentiation, enabling an end-to-end workflow from a new continuously varying multimaterial system to its mechanical simulation and design.} By extracting closed-form, composition-aware hyperelastic constitutive laws from \added{sparsified neural networks trained on} experimental data, this approach facilitates exact symbolic differentiation \deleted{via}\added{of the terms required by} the adjoint state method implemented with FEniCSx, \deleted{efficiently }circumventing the bottlenecks of applying neural network constitutive models\added{ in analysis and optimization}. This pipeline is deployed on soft robotic gripper applications, demonstrating continuous composition optimization for highly anisotropic contact responses, and the concurrent optimization of macroscopic topology and material distribution under non-failure stretch constraints. This \deleted{methodology could replace} \added{study is a step toward replacing conventional constitutive modeling and} laborious empirical prototyping, establishing interpretable machine-learning models as practical, robust design primitives for advanced multimaterial additive manufacturing.
  \end{abstract}


  %Research highlights
%   \begin{highlights}
%   \item Research highlight 1
%   \item Research highlight 2
%   \end{highlights}

\end{frontmatter}

% \pagewiselinenumbers

\section{Introduction}

Digital materials, in the context of multimaterial 3D printing, are blends of multiple base materials (usually polymeric) \cite{ahn20243DPrinting3D} that can be combined at different mixing ratios to realize spatially heterogeneous, functionally graded material distributions at the component level. Such spatial gradients are ubiquitous across various animal and plant species, from a squid's beak and a spruce trunk to bones \cite{rossetti2017microstructure} and lenses in human eyes \cite{pragya2023SoftFunctionallyGradient}, and have inspired advancements in soft robotics \cite{bartlett20153d}, aerospace engineering \cite{kumar2013development}, bioengineering \cite{saleh202030}, and beyond \cite{li2025high}, enabling target material performance and improved properties. Multimaterial 3D-printing techniques for digital materials, such as the PolyJet method, have achieved spatial resolutions as fine as 14 microns \cite{muthuram2022review,wei2019experimental}, allowing structures to be meticulously tailored for specific functionality \added{\cite{levin2023swelling}}. However, the combination of exploring component topology along with functionally graded material distribution creates a high-dimensional design space whose optimization is hindered by an additional bottleneck. A prerequisite for this design optimization problem is the efficient discovery and calibration of constitutive laws that capture the material response for a multitude of material combinations; the problem becomes more challenging as more base materials are utilized and mixed in continuously varying ratios. However, the conventional design paradigm for functionally graded materials relies heavily on empirical methods, characterized by a laborious trial-and-error process \cite{li2025high} common in prototyping. In contrast, numerical modeling and optimization offer the opportunity to design multimaterial distributions with significantly accelerated turnaround times \cite{wade2025implicit,smith2021seamless}, from material testing to prototyping. Nevertheless, the efficacy and efficiency of these numerical approaches are constrained by (1) the calibration of material models for multimaterial 3D printing, especially in cases involving continuous mixing ratios of base components, and (2) optimization strategies within a spatially high-dimensional design space. The multimaterial additive manufacturing community actively seeks to overcome these limitations \cite{pragya2023SoftFunctionallyGradient,li2025high,nazir2023multi,hirano2025modeling,garg2026machine} and calls for specialized computational tools 
\cite{nazir2023multi}.

Traditional constitutive modeling approaches, which involve phenomenologically proposing and empirically fitting formulations to a single material response \cite{bocciarelli2008constitutive,pascon2018large,xue2014phenomenological}, rely on workflows that are not automated and are heavily dependent on user expert-knowledge. In light of the universal approximation theorem of neural networks \cite{hornik1989multilayer,lu2019deeponet}, data-driven constitutive modeling provides a pathway toward such unified models \cite{fuhg2025ReviewDataDrivenConstitutive}, and has evolved significantly in recent years. Early data-hungry, model-free approaches \cite{kirchdoerferDatadrivenComputationalMechanics2016,tangMAP123DatadrivenApproach2019,luo2022data} directly replaced constitutive laws with experimental data or standard machine learning surrogate models \cite{guoDatadrivenApproachPredicting2022,klein2022polyconvex,xuLearningConstitutiveRelations2021}. However, when trained on scarce data, these pure data-driven models often suffer from overfitting and poor extrapolation. Consequently, the field has shifted toward physics-augmented methods \cite{linka2021constitutive,ciftciPhysicsinformedGANFramework2024,fuhg2022physics,fuhg2024stress,tac2022data,luo2025physics}, which embed prior physical knowledge into the data, network architecture, or training process to improve robustness, interpretability, and extrapolation capabilities. These physics-augmented models have been successfully applied to damage mechanics \cite{janssen2024physics}, multiphysics simulations \cite{luo2026physics}, and topology optimization \cite{garbrecht2021InterpretableMachineLearning}. By achieving high predictive accuracy for various material responses while reducing the computational burden of expensive multiscale or high-fidelity simulations \cite{xuLearningConstitutiveRelations2021}, they hold great potential for establishing unified constitutive models for functionally graded and digital materials, though this specific application remains an actively growing field \cite{garg2026machine}. The development and calibration of constitutive laws that recapitulate the response of digital materials become practically infeasible due to the combinatorial nature of mixing a large number of base materials in continuous mixing ratios. As such, physics-augmented machine learning approaches have proven to be a potential solution for this class of problems \cite{yangPhysicsAugmentedMachine2025}. 

Nonetheless, when applying computational frameworks such as the finite element method, standard machine-learning-based models lack explicit, closed-form constitutive expressions from which a Jacobian can be readily assembled. Consequently, evaluating the structural response typically requires forward-mode automatic differentiation \cite{revels2016forward} \added{and constructing appropriate preconditioners} to utilize Jacobian-free Newton-Krylov methods, while optimization tasks, particularly when involving asymmetric Jacobians, may rely on backward-mode automatic differentiation (as demonstrated in Section \ref{sec:adjoint-method-lagrangian-multiplier}). However, these Jacobian-free methods can result in increased computational time and degraded convergence for highly nonlinear or ill-conditioned problems \cite{knoll2004jacobian}, particularly when the material stability of data-driven constitutive models is not guaranteed. \added{Even with an assembled tangent, automatic differentiation of neural network constitutive models remains costly, with analytical derivatives reported to be 2--10 times faster \cite{alheit2026commet}.} In such scenarios, sparse-regression-based approaches that yield explicit symbolic expressions for constitutive laws \cite{brunton2016discovering} warrant greater attention, as computational frameworks like \texttt{FEniCSx} \cite{Baratta_DOLFINx_the_next_2023} favor explicit expressions for the efficient symbolic derivation of derivatives. Recent efforts have made significant progress in this relatively new direction \cite{flaschel2021unsupervised,fuhg2024extreme}, including recent work \cite{yangPhysicsAugmentedMachine2025} that specifically targets physics-augmented modeling to accurately represent and extrapolate in compositional space for digital materials\added{, although their potential in structural analysis and optimization has not been fully utilized}.

Effectively designing structures with digital materials requires not only accurate modeling but also optimization techniques for both topology and continuous material search spaces. Building upon traditional topology optimization, which seeks to optimally distribute material for maximum performance \cite{sigmund2013topology}, researchers have developed multi-material and multiscale optimization techniques \cite{yu2025multi,li2025multi,zhang2018multi}. Multimaterial topology optimization typically aims to distribute a discrete set of base materials by interpolating between multiple material phases with distinct properties. \added{In density-based formulations, one or more design variables per element (pseudo-densities) interpolate between void and each solid phase \cite{zuo2017multi,meng2025multi,sarkar2026generalized}, also for hyperelastic materials under large deformations \cite{zhang2020adaptive,song2026neo}. These schemes typically penalize intermediate values to recover a crisp selection among discrete materials, whereas functionally graded designs keep the local fraction of two constituents continuous \cite{silva2024minimizing} or grade the properties of a single printed material \cite{valizadeh2025minimizing}. Digital materials allow such grading, which avoids sharp transitions where high strain concentrations occur~\cite{saldivar2023bioinspired,bartlett20153d}. These interpolations rely on rules of mixtures, assumed energy interpolations, or empirical fits, which are rarely calibrated for the nonlinear, composition-dependent hyperelastic response of digital materials under finite deformation; only recently has this response been characterized and modeled \cite{yangPhysicsAugmentedMachine2025}.} Alternatively, multiscale topology optimization expands the design space by concurrently designing microscale architectures and their macroscopic arrangement \cite{conde2024multi,guo2025multi,shimoda2024concurrent}. While this approach is well-suited to the hierarchical nature of functionally graded \added{materials} and \added{the need for smooth transitions of} digital materials, the additional scale of simulation introduces significant computational costs \cite{wu2021topology,ma2024asymptotic} and substantial implementation effort. The archetype of these problems directly maps to the concurrent topology and material distribution optimization problem for multimaterial 3D printing\added{, where a composition ratio $c$ describes the blend of base materials and the response at intermediate compositions is learned from tests \cite{yangPhysicsAugmentedMachine2025}}. \added{More fundamentally, conventional optimization relies on deriving the sensitivities, a laborious step that needs to be repeated for multiple constitutive models, objectives, or constraints, especially complex ones. }

Recent advances in multiscale topology optimization have addressed its high-dimensional and computationally expensive nature by utilizing surrogate, homogenized constitutive laws. These laws often employ a single parameter to represent microscopic structural variations, which is then treated as a macroscopic field variable during optimization. This parameter field can be optimized independently \cite{ituarte2019DesignAdditiveManufacture} or concurrently with the macroscopic topology \cite{vijayakumaran2025ConsistentMachineLearninga,wei2026MachineLearningbasedMultiscale,stollberg2025multiscale}. While it is possible to employ a simple empirical property-variable relationship \cite{ituarte2019DesignAdditiveManufacture}, most recent approaches incorporate machine-learning-based constitutive laws to accelerate the process and ideally enable scaling to more complex scenarios. For instance, \deleted{Seo and Min \cite{seo2022novel} utilized deep generative and convolutional regression models to map latent representations to microstructural images and subsequently to linear elastic properties.}\added{Stollberg et al. \cite{stollberg2025multiscale} used a physics-augmented neural network to map the parameters of functionally graded lattices to their linear elastic stiffness, whose sensitivities were obtained by automatic differentiation of the network.} Vijayakumaran et al. \cite{vijayakumaran2025ConsistentMachineLearninga} trained an Input Convex Neural Network (ICNN) to \deleted{directly map strain to stress}\added{represent the strain energy density} for hyperelastic materials with varying microstructures. However, their approach did not fully resolve the computationally expensive neural network differentiation required for\added{ both} the forward and \deleted{inverse processes} \added{sensitivity analyses} essential for scaling these approaches. \added{Jadoon et al. \cite{jadoon2026multiscale} avoided automatic differentiation by hand-deriving the derivatives of a dense physics-augmented network, an effort that must be repeated whenever the network, objective, or constraints change.} More recently, Wei et al. \cite{wei2026MachineLearningbasedMultiscale} highlighted the importance of explicit constitutive expressions to facilitate simulation and optimization. They used symbolic regression to extract expressions from multiple ICNNs, although this supplementary step introduces additional implementation complexity and a potential loss of accuracy\added{, and still leaves the design sensitivities to automatic differentiation}, contrary to the scheme for extreme sparsification of \deleted{PANNs}\added{physics-augmented neural networks (PANNs)} proposed by Fuhg et al. \cite{fuhg2024extreme}\added{, in which the trained network itself is the closed-form expression}. Sparsification leads to a direct pathway for the integration of such \deleted{physics-augmented neural network (PANN)}\added{PANN} constitutive laws directly into finite element solvers, as recently showcased for multi-modal and multi-fidelity constitutive model development from digital image correlation data and simple experiments \cite{tan2026towards}.

\deleted{This work aims to provide an end-to-end pipeline for designing 3D-printed multimaterial structures by combining: (1) sparse regression to learn explicit constitutive laws for digital materials using PANNs, (2) leveraging the symbolic-differentiation-based finite element framework \texttt{FEniCSx} to promote efficient numerical simulations involving PANN-based constitutive laws, and (3) the concurrent optimization of their topology and material distribution. This workflow emphasizes the use of scarce, low-cost material test data and the derivation-free implementation of constitutive models and optimization definitions, alleviating the burden of laborious empirical trial-and-error processes.} \added{To make machine-learning constitutive laws a practical replacement for conventional constitutive modeling in analysis and design, this work puts forward an end-to-end pipeline from a new continuously varying multimaterial system and its scarce, low-cost material test data, to neural network constitutive laws, and to finite element analysis and optimization with complex objectives and constraints. Sparse regression of PANNs reduces the trained network to a compact closed-form law, from which the symbolic-differentiation-based finite element framework \texttt{FEniCSx} generates the tangent and the sensitivities of the objectives and constraints, without the runtime automatic differentiation through a network or the hand-derived constitutive derivatives that previous studies relied on, and without a bespoke software interface. The pipeline is demonstrated by the concurrent optimization of topology and material distribution for a 3D-printed polymeric multimaterial system.} The remainder of the paper is organized as follows: Section \ref{sec:methodology} details the PANN-based constitutive law (Section \ref{sec:ml-constitutive}) and optimization techniques, including the adjoint state method (Section \ref{sec:adjoint-method-lagrangian-multiplier}), topology optimization (Section \ref{sec:simp}), and numerical implementation (Sections \ref{sec:filter} and \ref{sec:implementation}). Section \ref{sec:results} focuses on applying this workflow to problems in the design of robotic gripper components, specifically exploring the anisotropic response of contact points on robotic hands under small (Section \ref{sec:contact-point-small-deform}) and large (Section \ref{sec:contact-points-large-deform}) deformations, alongside the damage-aware optimization of robotic gripper fingers (Section \ref{sec:clamping-fingers}). Finally, Section \ref{sec:conclusions} summarizes the findings and discusses limitations and avenues for future development.  

\section{Theory and implementation}
\label{sec:methodology}

This section details our design workflow, beginning with a composition-aware, machine-learning-based constitutive law for digital materials (Section \ref{sec:ml-constitutive}). This is followed by the optimization strategy for topology and material composition using the adjoint state method (Section \ref{sec:adjoint-method-lagrangian-multiplier} and Section \ref{sec:simp}), the filtering of optimized field variables (Section \ref{sec:filter}), and other key implementation aspects (Section \ref{sec:implementation}). The section focuses on the necessary theory for each of the components of the workflow, but also details the implementation choices required for numerical experiments.

\subsection{Machine-learning-based constitutive models for digital material}
\label{sec:ml-constitutive}

\subsubsection{PANNs for digital material}

Recent work of the authors \cite{yangPhysicsAugmentedMachine2025} proposed a compact, closed-form constitutive law for digital materials using a partial Input Convex Neural Network (pICNN) calibrated from experimental data. The model specifically focused on blends of Agilus with \deleted{a Digital ABS (}RGD-531\deleted{)} at different ratios (termed composition ratio\deleted{ $c$}), given a fixed percentage of \deleted{another Digital ABS (}RGD-515\deleted{)}, printed by a Stratasys Objet260 Connex3 printer. \added{Among the three blended materials, Agilus is a flexible photopolymer with high tear resistance and tensile strength (relative to other flexible photopolymers), the ability to withstand repeated flexing and bending cycles, great elongation at break (around 220\%--240\%), and low Shore hardness; RGD-531 and RGD-515 are rigid photopolymers that together form Digital ABS, which simulates standard ABS plastics with high stiffness (elastic modulus of 2.6--3.0~GPa), thermal resistance, and impact toughness. Conversely, Agilus has a low stiffness and tensile strength (2.4--3.1~MPa), whereas Digital ABS has a limited elongation at break (25--40\%), so that neither base material alone combines stiffness with stretchability. In this work, such a blend of these materials facilitates the design of an effective anisotropic structure by utilizing their distinct stiffnesses, and the design of a lightweight stiff structure that also tolerates local strain concentration by utilizing the great elongation at break of Agilus. }

Although the original model incorporated both incompressible hyperelasticity and viscoelasticity, the present work focuses exclusively on the hyperelastic quasi-static response with a vision towards integration in design optimization workflows. The original model was trained \added{and tested} on 30 sets of uniaxial tension and torsion experiments conducted at three loading rates across five composition ratios. \added{The composition ratio $c$ represents the Digital ABS (RGD-515 and RGD-531) content relative to Agilus, scaled such that $c=0$ and $c=1$ correspond to the blends with 24.8\% and 27.2\% Digital ABS by mass, respectively, while $c>1$ denotes a higher Digital ABS content (e.g., $c=2.0895$ for 29.7\%) \cite{yangPhysicsAugmentedMachine2025}. Three of the five composition ratios, normalized to $c=0$, 0.4669, and 1.0, are for training, while the other two composition ratios $c=0.1755$ and 2.0895 are held out for testing. The model accurately predicts the held-out composition $c=0.1755$ within the training range ($R^2\geq0.96$), whereas its accuracy deteriorates for the extrapolated composition $c=2.0895$ \cite{yangPhysicsAugmentedMachine2025}. } If focusing on the hyperelastic behavior, the requisite \deleted{training} data could have been reduced to just 10 sets of curves. The work in \cite{yangPhysicsAugmentedMachine2025} showcased the concurrent discovery and calibration of the pICNN-based constitutive laws, but did not focus on FEM deployment for further optimization tasks. Directly from \cite{yangPhysicsAugmentedMachine2025}, the incompressible hyperelastic strain energy density is expressed as
\begin{equation}
    \psi(\mathbf{u},c)=\psi^{NN}\left(\bar{I}_1,\bar{I}_2,c\right)-\underset{\mathrm{Normalization}}{\underbrace{\psi^{NN}(3,3,c)}}+\underset{\mathrm{Incompressibility\ hard\ constraint}}{\underbrace{p \left(J-1\right)}},
    \label{eq:total_strain_energy_NN}
\end{equation}
where $\psi^{NN}$ is the learned expression detailed subsequently. The modified strain invariants are $\bar{I}_1=\mathrm{tr}\bar{\mathbf{C}}$ and $\bar{I}_2=\frac{1}{2}\left[\left(\mathrm{tr} \bar{\mathbf{C}}\right)^2-\mathrm{tr}\left(\bar{\mathbf{C}}^2\right)\right]$, derived from the isochoric right Cauchy-Green tensor $\bar{\mathbf{C}}=J^{-2/3}\mathbf{C}$, which decouples the isochoric deformation from the standard right Cauchy-Green tensor $\mathbf{C}=\mathbf{F}^\mathrm{T}\mathbf{F}$. Here, $\mathbf{F}$ represents the deformation gradient, $J = \det(\mathbf{F})$, and $p$ serves as a Lagrange multiplier field representing hydrostatic pressure to enforce exact incompressibility.

The strain energy density $\psi^{NN}$ is represented by a pICNN, which \deleted{combines an input convex neural network (ICNN) with non-convex and fully convex (NCFC) layers, alongside purely non-convex (NC) layers} \added{consists of fully convex (FC) layers forming an input convex neural network (ICNN), purely non-convex (NC) layers, and non-convex-to-fully-convex (NCFC) layers connecting the two} \cite{yangPhysicsAugmentedMachine2025}. \added{The architecture is designed so that $\psi^{NN}$ is convex with respect to the strain invariants $\bar{\mathbf{I}}=[\bar{I}_1,\bar{I}_2]^\mathrm{T}$, while its dependence on the composition ratio $c$ is left unrestricted. All hidden layers are activated by the $\mathrm{Softplus}$ function $\mathrm{Softplus}(x)=\ln(1+e^{x})$, which is smooth, convex, and monotonically non-decreasing. First, the NC layers process the composition ratio through a sequence of fully connected layers:}
\begin{equation}
\added{\mathbf{y}_1=\mathrm{Softplus}\left(\mathbf{W}_{nc}^{(0)}c\right),\quad \mathbf{y}_h=\mathrm{Softplus}\left(\mathbf{W}_{nc}^{(h-1)}\mathbf{y}_{h-1}\right),\quad h=2,\dots,H,}
\label{eq:picnn-nc}
\end{equation}
\added{where $H$ is the number of hidden layers and $\mathbf{y}_h$ are the composition features. No sign constraint is imposed on the weights $\mathbf{W}_{nc}^{(h)}$, so that the NC layers can capture the nonlinear and possibly non-monotonic influence of composition on the mechanical response, for which convexity is not physically required. Second, the NCFC layers map the composition features into $\mathbf{c}_h=\mathrm{Softplus}(\mathbf{W}_{ncfc}^{(h-1)}\mathbf{y}_h)$, so that the composition nonlinearly modulates the strain energy at every hidden layer. Third, the FC layers form the ICNN path~\cite{amos2017input}, which maps the strain invariants to the strain energy, taking $\mathbf{c}_h$ from the NCFC layers as additional inputs:}
\begin{equation}
\begin{aligned}
\added{\mathbf{x}_1}&\added{=\mathrm{Softplus}\left(\mathbf{W}_{I}^{(0)}\bar{\mathbf{I}}+\mathbf{W}_{c}^{(0)}\mathbf{c}_1\right),}\\
\added{\mathbf{x}_h}&\added{=\mathrm{Softplus}\left(\mathbf{W}_{I}^{(h-1)}\bar{\mathbf{I}}+\mathbf{W}_{x}^{(h-1)}\mathbf{x}_{h-1}+\mathbf{W}_{c}^{(h-1)}\mathbf{c}_h\right),\quad h=2,\dots,H,}\\
\added{\psi^{NN}}&\added{=\mathbf{W}_{x}^{(H)}\mathbf{x}_H+\mathbf{W}_{I}^{(H)}\bar{\mathbf{I}}+\mathbf{W}_{c}^{(H)}\mathbf{c}_H.}
\end{aligned}
\label{eq:picnn-fc}
\end{equation}
\added{The weights $\mathbf{W}_{I}^{(0)}$ acting on the invariants in the first layer and all the weights $\mathbf{W}_{c}^{(h)}$ acting on the composition features are unconstrained, whereas all the weights acting on the hidden states ($\mathbf{W}_{x}^{(h)}$) and the invariant weights of the subsequent layers ($\mathbf{W}_{I}^{(h)}$, $h\geq1$) are constrained to be non-negative. Since $\mathbf{c}_h$ is independent of $\bar{\mathbf{I}}$, each component of $\mathbf{x}_1$ is convex in $\bar{\mathbf{I}}$, and non-negative combinations of convex functions passed through the convex, non-decreasing $\mathrm{Softplus}$ preserve convexity layer by layer. Hence, $\psi^{NN}$ is convex in $\bar{I}_1$ and $\bar{I}_2$ for any composition $c$. Convexity in the invariants is a key ingredient of polyconvexity, which guarantees material stability and the existence of minimizers of the boundary value problems~\cite{yangPhysicsAugmentedMachine2025}. This is essential for robust finite element solutions throughout the optimization, where $c$ varies continuously within the design domain. The trainable parameters are grouped as $\boldsymbol{\theta}=[\boldsymbol{\theta}_{fc},\boldsymbol{\theta}_{nc},\boldsymbol{\theta}_{ncfc}]$, collecting the weights of the FC, NC, and NCFC layers, respectively.}

\added{The network was calibrated against uniaxial tension and torsion experiments. For uniaxial tension, the incompressible kinematics and the traction-free lateral surfaces determine the predicted axial Cauchy stress $\hat{\sigma}_{11}$ from the stretch, while for torsion of a cylindrical rod, the predicted torque $\hat{T}$ is obtained by integrating the shear stress $\sigma_{z\theta}$ derived from $\psi^{NN}$ over the cross-section~\cite{yangPhysicsAugmentedMachine2025}. For the $k$-th tension or torsion experiment, the predictions at all recorded load steps are collected in the vectors $\hat{\boldsymbol{\sigma}}_{11}^{(k)}$ and $\hat{\mathbf{T}}^{(k)}$, and the corresponding measurements in $\boldsymbol{\sigma}_{11}^{(k)}$ and $\mathbf{T}^{(k)}$. The loss function reads}
\begin{equation}
\begin{aligned}
\added{\mathcal{E}(\boldsymbol{\theta})}&\added{=\frac{w_{\mathrm{ten}}}{N_{\mathrm{ten}}}\sum_{k=1}^{N_{\mathrm{ten}}}\left\|\frac{\boldsymbol{\sigma}_{11}^{(k)}-\hat{\boldsymbol{\sigma}}_{11}^{(k)}}{\max\left(\boldsymbol{\sigma}_{11}^{(k)}\right)-\min\left(\boldsymbol{\sigma}_{11}^{(k)}\right)}\right\|_2^2+\frac{w_{\mathrm{tor}}}{N_{\mathrm{tor}}}\sum_{k=1}^{N_{\mathrm{tor}}}\left\|\frac{\mathbf{T}^{(k)}-\hat{\mathbf{T}}^{(k)}}{\max\left(\mathbf{T}^{(k)}\right)-\min\left(\mathbf{T}^{(k)}\right)}\right\|_2^2}\\
&\added{\quad+\alpha_{fc}\left\|\boldsymbol{\theta}_{fc}\right\|_0+\alpha_{nc}\left\|\boldsymbol{\theta}_{nc}\right\|_0+\alpha_{ncfc}\left\|\boldsymbol{\theta}_{ncfc}\right\|_0,}
\end{aligned}
\label{eq:picnn-loss}
\end{equation}
\added{where $N_{\mathrm{ten}}$ and $N_{\mathrm{tor}}$ are the numbers of tension and torsion experiments. The first two terms are range-normalized squared errors, summed over the recorded load steps and averaged over the experiments, that measure the data fidelity, with weights $w_{\mathrm{ten}}$ and $w_{\mathrm{tor}}$ balancing the two types of tests. The last three terms are $L_0$ penalties, which count the non-zero parameters in each group, weighted by the coefficients $\alpha_{fc}$, $\alpha_{nc}$, and $\alpha_{ncfc}$. The $L_0$ penalty drives redundant parameters exactly to zero without shrinking the remaining ones, enabling a compact closed-form expression. Since the $L_0$ norm is not differentiable, it is relaxed following Louizos et al.~\cite{louizos2017learning}: each parameter in $\boldsymbol{\theta}$ is multiplied by a stochastic gate in $[0,1]$ whose distribution is controlled by a trainable parameter, the $L_0$ norm in the loss function is replaced by the expected number of open gates, and the data-fidelity terms are averaged over Monte Carlo samples of the gates. After training, the gates are evaluated deterministically.}

\added{The pICNN was trained with the Adam optimizer~\cite{kingma2014adam}, starting from an over-parameterized architecture with two FC layers of 30 neurons and two NC layers of 5 neurons (over 1,400 weights in total). Since the original model also captured rate dependence through a quasi-linear viscoelastic model~\cite{yangPhysicsAugmentedMachine2025}, the training was staged to introduce one effect at a time: the pICNN was first fitted to the instantaneous elastic response using the tension tests at the fastest stretch rate; the viscoelastic parameters and the pICNN were then refined alternately using the tension tests at all rates; the torsion data were added next; and finally, the $L_0$ penalties were gradually activated. This final stage pruned the vast majority of the weights without degrading the fit to the experimental data. Only the resulting hyperelastic potential $\psi^{NN}$ is used in the present work, since the optimization problems considered herein are quasi-static.} \deleted{Owing to an $L_0$ regularization term in the loss function, which strictly penalizes the number of non-zero parameters in the network, $\psi^{NN}$ can} \added{After pruning, $\psi^{NN}$ can} be explicitly formulated using only 21 parameters:
\begin{equation}
\begin{aligned}
    \psi^{NN}&={0.083089}\bar{I}_1+{0.809351}\ln\left( 1+e^{{0.009634}\bar{I}_1} \right) \\
    &+ {4.01198}\ln\left\{ 1+e^{{0.033506}\bar{I}_1}
    \left[1+e^{{-0.207897}\bar{I}_1+{0.135303}\bar{I}_2}\left( 1+\omega\right)^{{-0.021823}}\right]^{{1.16891}}\right.\\
    &\quad\quad\quad\quad\quad\quad\quad\quad\quad\quad\quad\quad\left.\left[1+e^{{0.228289}\bar{I}_1{-0.204157}\bar{I}_2}\left( 1+\omega\right)^{{0.049792}}\right]^{{0.733048}}
    \right\},
    \label{eq:psi-NN}
\end{aligned}
\end{equation}
where 
\begin{equation}
\omega=\left(1+e^{{6.84602}c}\right)^{{2.37126}}\left(1+e^{{6.86996}c}\right)^{{2.19610}}\left(1+e^{{6.88268}c}\right)^{{2.56167}}\left(1+e^{{6.93876}c}\right)^{{1.98880}}.
\end{equation}
The terms involving $e^{(\cdot)}$, $\ln(\cdot)$, and $1+(\cdot)$ originate from $\mathrm{Softplus}$ activation functions. This explicit expression enables the exact symbolic derivation of the stress and tangent stiffness tensors. Consequently, it eliminates the need for algorithmic workarounds to accelerate solvers or reliance on computationally expensive Jacobian-free methods that frequently suffer from poor convergence.

The total strain energy density $\psi$ is normalized by the second term on the right side of Eq.~\eqref{eq:total_strain_energy_NN} to ensure zero initial stress in the undeformed state (i.e., $\mathbf{C}=\mathbf{I}$). The third term introduces a kinematic constraint to enforce exact incompressibility ($J=1$), consequently ensuring that the standard invariants of $\mathbf{C}$ equal their isochoric counterparts ($I_1=\bar{I}_1$ and $I_2=\bar{I}_2$). As previously established, $p$ serves as the Lagrange multiplier for this constraint. To satisfy the stationary Karush-Kuhn-Tucker (KKT) conditions, $p$ is treated as an independent primary variable and solved concurrently with the displacement field via a mixed finite element approach utilizing Taylor-Hood elements. From this strain energy, the second Piola-Kirchhoff (PK2) stress is analytically computed as $\mathbf{S}=2\partial \psi/\partial \mathbf{C}$, which then yields the Cauchy stress $\boldsymbol{\sigma}=J^{-1}\mathbf{F}\mathbf{S}\mathbf{F}^\mathrm{T}$. These derived stress measures are subsequently utilized to train the parameters of $\psi^{NN}$ against experimental data. 

\subsubsection{Simplification to nearly incompressible model}
\label{sec:simp-nearly-incompressible}

Strictly enforcing incompressibility via an independent hydrostatic pressure field often introduces severe numerical instabilities, such as volumetric locking or spurious pressure oscillations, particularly under complex geometric and boundary conditions. As the aim of the work here is to enable topology and material distribution optimization in complex boundary value problems, a penalty approach is utilized for near incompressibility instead of a fully incompressible scheme. It is noted (as discussed in \cite{yangPhysicsAugmentedMachine2025}) that volumetric data were inconclusive from the torsional and uniaxial experiments; as such, exact incompressibility was also an approximation. The penalty term modifies the original expression to:
\begin{equation}
    \psi=\psi^{NN}\left(\bar{I}_1,\bar{I}_2,c\right)-\underset{\mathrm{Normalization}}{\underbrace{\psi^{NN}(3,3,c)}}+\underset{\mathrm{Incompressibility\ soft\ constraint}}{\underbrace{\frac{1}{2}\kappa \left(J-1\right)^2}},
    \label{eq:total_strain_energy_NN_soft}
\end{equation}
where the bulk modulus $\kappa$ acts as the penalty parameter to be calibrated. To calibrate this parameter, \texttt{sympy} is utilized to symbolically derive the relationship between the first Piola-Kirchhoff (PK1) stress and stretch in the infinitesimal strain regime, and evaluate it at a stretch of 1.001 to provide an estimate of the tensile modulus $E$ across various composition ratios $c$. \added{This is performed within the range $[0,1]$ where the pICNN is trained and can interpolate well ($R^2\geq0.96$)~\cite{yangPhysicsAugmentedMachine2025}.} Subsequently, the relationship between $E$ and $c$ is calibrated through a least-squares regression (Fig.~\ref{fig:fit_tensile_modulus}). The fitted relation is expressed as:
\begin{equation}
    E(c)=0.304c^4-0.962c^3+0.899c^2+0.399c+0.697,
    \label{eq:fitted-E-c}
\end{equation}
and the corresponding bulk modulus is calculated via standard isotropic elasticity relations as:
\begin{equation}
    \kappa=\frac{E(c)}{3(1-2\nu)},
\end{equation}
where Poisson's ratio $\nu$ is set to $0.49$ to accurately reflect a nearly-incompressible material response.

\subsection{Adjoint state method}
\label{sec:adjoint-method-lagrangian-multiplier}

The explicit strain energy density function derived via the pICNN facilitates mechanical simulations (i.e., the forward problem) for newly designed digital material systems. Complementing this forward analysis, the adjoint state method introduced in this section enables the efficient gradient computation of an objective function and structural constraints with respect to the spatial material distribution \added{(i.e., sensitivity analysis)}, which is essential for the \deleted{inverse problem of} \added{structural} design optimization\added{, with as many design variables as the finite element discretization admits and far fewer objective functions and constraints}.

Consider the following constrained minimization problem:
\begin{equation}
\begin{aligned}
    \min_v\ &\mathcal{J}(\mathbf{u},v), \\
    \mathrm{s.t.\ } &\text{\deleted{$\displaystyle f(\mathbf{u};v)=0,$}}\delcr\delamp\added{f(\mathbf{u},v,\delta\mathbf{u})=0\quad\forall\,\delta\mathbf{u}\in\mathcal{V}_h,}
\end{aligned}
\label{eq:constrained_min}
\end{equation}
\deleted{where $\mathbf{u}$ is the state variable governed by the state equation $f(\mathbf{u};v)=0$, and $v$ represents the design variable field.} \added{where the state variable (displacement) field $\mathbf{u}\in\mathcal{U}_h$ and the design variable field $v\in\mathcal{D}_h$ are defined in finite element spaces. Here, $\mathcal{U}_h$ contains the displacement fields satisfying the prescribed Dirichlet boundary conditions, and $\mathcal{V}_h$ is the corresponding space of test functions (virtual displacements) $\delta\mathbf{u}$ that vanish on the Dirichlet boundary. The state equation is the weak form of equilibrium, whose residual form $f(\mathbf{u},v,\delta\mathbf{u})$ is nonlinear in $\mathbf{u}$ and $v$ but linear in the test function $\delta\mathbf{u}$. The objective functional is considered in the form $\mathcal{J}(\mathbf{u},v)=\int_\Omega j(\mathbf{u},\nabla\mathbf{u},v)\,\mathrm{d}V+\int_{\partial\Omega}j_{\partial}(\mathbf{u},\nabla\mathbf{u},v)\,\mathrm{d}A$, where the densities $j$ and $j_{\partial}$ may depend on both the displacement and its gradient.} For simplicity, $v$ is denoted without boldface, although it may represent either a scalar or a vector field. \deleted{Computing the total derivative of $\mathcal{J}$ with respect to $v$ requires the sensitivity of the state variable ($\partial \mathbf{u}/\partial v$).} \added{Computing the total derivative of $\mathcal{J}$ with respect to $v$ requires the sensitivity of the state variable, i.e., $\mathrm{d}\mathbf{U}/\mathrm{d}\mathbf{v}$ for the nodal vectors $\mathbf{U}$ and $\mathbf{v}$ of $\mathbf{u}$ and $v$ defined below.} Evaluating this sensitivity via finite differences is computationally prohibitive for high-dimensional design spaces; the adjoint state method circumvents this bottleneck. \deleted{By introducing a Lagrange multiplier $\boldsymbol{\lambda}$ (the adjoint state variable), the Lagrangian functional is defined as} \added{By introducing a Lagrange multiplier $\boldsymbol{\lambda}\in\mathcal{V}_h$ (the adjoint state variable), which takes the place of the test function in the state equation, the Lagrangian functional is defined as}
\begin{equation}
\begin{aligned}
    &\text{\deleted{$\displaystyle \mathcal{L}(\mathbf{u},v,\boldsymbol{\lambda})=\mathcal{J}(\mathbf{u},v)+\left\langle f(\mathbf{u};v),\boldsymbol{\lambda} \right\rangle,$}}\delcr
    \delamp\added{\mathcal{L}(\mathbf{u},v,\boldsymbol{\lambda})=\mathcal{J}(\mathbf{u},v)+f(\mathbf{u},v,\boldsymbol{\lambda}).}
\end{aligned}
\end{equation}
\deleted{where $\langle\cdot,\cdot\rangle$ denotes the inner product.} \added{Throughout, test functions enter forms as ordinary (comma-separated) arguments, whereas the direction of a Gâteaux derivative is written after a semicolon:}
\begin{equation}
\added{D_y \Phi(y,w;\delta y)=\left.\frac{\mathrm{d}}{\mathrm{d}\epsilon}\Phi(y+\epsilon\delta y,w)\right|_{\epsilon=0},}
\label{eq:gateaux}
\end{equation}
\added{which denotes the Gâteaux derivative of a functional $\Phi$ with respect to $y$ in the direction $\delta y$ and is linear in $\delta y$, where $w$ collects the remaining arguments, including the other fields and the test function.} The optimal solution requires the stationarity of $\mathcal{L}$ with respect to $\mathbf{u}$, $v$, and $\boldsymbol{\lambda}$, meaning all directional derivatives must vanish\added{ for arbitrary admissible directions}: 
\begin{equation}
\begin{aligned}
\text{\deleted{$\displaystyle D_\mathbf{u}\mathcal{L}$}}\delamp\text{\deleted{$\displaystyle \mathord{}=D_\mathbf{u}\mathcal{J}+\left\langle D_\mathbf{u} f(\mathbf{u};v),\boldsymbol{\lambda} \right\rangle$}}\delcr
\delamp\text{\deleted{$\displaystyle \mathord{}=\left\langle\nabla_\mathbf{u} \mathcal{J},\delta_\mathbf{u}\right\rangle+\left\langle \nabla_\mathbf{u} f(\mathbf{u};v)\delta_\mathbf{u},\boldsymbol{\lambda} \right\rangle=0,$}}\delcr
\text{\deleted{$\displaystyle D_v\mathcal{L}$}}\delamp\text{\deleted{$\displaystyle \mathord{}=D_v\mathcal{J}+\left\langle D_vf(\mathbf{u};v),\boldsymbol{\lambda} \right\rangle=0,$}}\delcr
\text{\deleted{$\displaystyle D_{\boldsymbol{\lambda}} \mathcal{L}$}}\delamp\text{\deleted{$\displaystyle \mathord{}=f(\mathbf{u};v)=0,$}}\delcr
\added{D_\mathbf{u}\mathcal{L}(\mathbf{u},v,\boldsymbol{\lambda};\delta\mathbf{u})}&\added{\mathord{}=D_\mathbf{u}\mathcal{J}(\mathbf{u},v;\delta\mathbf{u})+D_\mathbf{u}f(\mathbf{u},v,\boldsymbol{\lambda};\delta\mathbf{u})=0\quad\forall\,\delta\mathbf{u}\in\mathcal{V}_h,} \\
\added{D_v\mathcal{L}(\mathbf{u},v,\boldsymbol{\lambda};\delta v)}&\added{\mathord{}=D_v\mathcal{J}(\mathbf{u},v;\delta v)+D_vf(\mathbf{u},v,\boldsymbol{\lambda};\delta v)=0\quad\forall\,\delta v\in\mathcal{D}_h,} \\
\added{D_{\boldsymbol{\lambda}}\mathcal{L}(\mathbf{u},v,\boldsymbol{\lambda};\delta\boldsymbol{\lambda})}&\added{\mathord{}=f(\mathbf{u},v,\delta\boldsymbol{\lambda})=0\quad\forall\,\delta\boldsymbol{\lambda}\in\mathcal{V}_h.}
\end{aligned}
\label{eq:lagrange_solution}
\end{equation}
\deleted{Here, $D_y f(x)=\left.\frac{\mathrm{d}}{\mathrm{d}\epsilon}f(x+\epsilon\delta_y)\right|_{\epsilon=0}$ denotes the Gâteaux derivative with respect to $y$, where $\delta_y$ represents an arbitrary variation in the space of $y$. The identity $D_y f(x)=\langle \nabla_y f(x), \delta_y \rangle$ is also utilized. While $D_{v} \mathcal{L}$ represents the exact derivative of the continuous Lagrangian, numerical optimizers require a discrete array of sensitivities. Therefore, after projecting the continuous derivative onto the finite-dimensional finite element space, the notation $\nabla_{\mathbf{v}} \mathcal{L}$ explicitly denotes the resulting column vector of nodal gradients, where $\mathbf{v}$ is the column vector of nodal design variables.} \added{Because a variation of the displacement also varies its gradient, the derivative of the objective functional contains contributions from both $\mathbf{u}$ and $\nabla\mathbf{u}$:}
\begin{equation}
\begin{aligned}
\added{D_\mathbf{u}\mathcal{J}(\mathbf{u},v;\delta\mathbf{u})}&\added{\mathord{}=\int_\Omega\left(\frac{\partial j}{\partial\mathbf{u}}\cdot\delta\mathbf{u}+\frac{\partial j}{\partial\nabla\mathbf{u}}:\nabla\delta\mathbf{u}\right)\mathrm{d}V}\\
&\added{\quad+\int_{\partial\Omega}\left(\frac{\partial j_{\partial}}{\partial\mathbf{u}}\cdot\delta\mathbf{u}+\frac{\partial j_{\partial}}{\partial\nabla\mathbf{u}}:\nabla\delta\mathbf{u}\right)\mathrm{d}A,}
\end{aligned}
\label{eq:objective_variation}
\end{equation}
\added{and the derivatives of $f$ are obtained in the same manner. Since all fields and test functions lie in finite element spaces, $\mathcal{J}$ and $f$ are functions of the nodal vectors of these fields, and their Gâteaux derivatives in the directions of basis functions are partial derivatives with respect to nodal values, as used in the three steps below. Eq.~\eqref{eq:lagrange_solution} is therefore the optimality system of the discretized problem, i.e., the sensitivities follow a discretize-then-differentiate approach~\cite{fernandez2018semi}, consistent with the symbolic differentiation of the discretized UFL forms in \texttt{FEniCSx} (Section~\ref{sec:implementation}). }

The adjoint state method consists of three steps: 
\begin{enumerate}
    \item (The forward problem) \deleted{Eq.~\eqref{eq:lagrange_solution}c is solved for the state variable $\mathbf{u}$, yielding the solution $\tilde{\mathbf{u}}$.} \added{Writing $\mathbf{u}=\mathbf{u}_{\mathrm{D}}+\sum_j U_j\mathbf{N}_j$, where $\mathbf{u}_{\mathrm{D}}$ satisfies the prescribed Dirichlet boundary conditions and $\mathbf{U}$ is the nodal vector of the free displacement degrees of freedom, and substituting the basis functions $\mathbf{N}_i$ of $\mathcal{V}_h$ as test functions, Eq.~\eqref{eq:lagrange_solution}c becomes the nonlinear system $f(\mathbf{u},v,\mathbf{N}_i)=0$ for all $i$, which is solved for $\mathbf{U}$, yielding the solution $\tilde{\mathbf{U}}$.} To achieve this, if the explicit expression of $f$ is known, the global Jacobian matrix $\mathbf{J}$\deleted{$\mathord{}=\nabla_\mathbf{u} f(\mathbf{u};v)$}\added{ with entries $(\mathbf{J})_{ij}=D_\mathbf{u}f(\mathbf{u},v,\mathbf{N}_i;\mathbf{N}_j)=\partial f(\mathbf{u},v,\mathbf{N}_i)/\partial U_j$} is typically assembled, and a Newton-Raphson scheme is employed with either a direct or a Newton-Krylov iterative solver. Alternatively, a Jacobian-free Newton-Krylov (JFNK) iterative solver can be utilized if the Jacobian-vector product (JVP) is directly accessible via forward-mode automatic differentiation. This latter approach is common for complex implicit formulations or standard machine-learning-based constitutive laws \cite{revels2016forward}. However, in the Jacobian-free case, the lack of an explicitly assembled Jacobian matrix hinders the construction of effective preconditioners, often leading to degraded convergence for highly nonlinear or ill-conditioned systems \cite{knoll2004jacobian}.  
    \item (The adjoint problem) \deleted{Eq.~\eqref{eq:lagrange_solution}a forms a linear system of equations that is solved for the adjoint variable $\boldsymbol{\lambda}$, yielding the solution $\tilde{\boldsymbol{\lambda}}$.} \added{Because $f$ is linear in its test function argument, writing $\boldsymbol{\lambda}=\sum_i\Lambda_i\mathbf{N}_i$ and substituting $\delta\mathbf{u}=\mathbf{N}_j$ turns Eq.~\eqref{eq:lagrange_solution}a into the linear system $\mathbf{J}^\mathrm{T}\boldsymbol{\Lambda}=-\nabla_\mathbf{U}\mathcal{J}$, with $(\nabla_\mathbf{U}\mathcal{J})_j=D_\mathbf{u}\mathcal{J}(\mathbf{u},v;\mathbf{N}_j)=\partial\mathcal{J}/\partial U_j$, which is solved for the nodal vector $\boldsymbol{\Lambda}$ of the adjoint variable, yielding the solution $\tilde{\boldsymbol{\Lambda}}$.} %This step can be viewed as the application of the Implicit Function Theorem. 
    \deleted{Solving the weak form in Eq.~\eqref{eq:lagrange_solution}a requires isolating the arbitrary variation $\delta_\mathbf{u}$ within the inner product. This is achieved by introducing an adjoint operator $(\cdot)^*$ such that $\left\langle \nabla_\mathbf{u} f(\mathbf{u};v)\delta_\mathbf{u},\boldsymbol{\lambda} \right\rangle=\left\langle \delta_\mathbf{u},(\nabla_\mathbf{u} f(\mathbf{u};v))^*\boldsymbol{\lambda} \right\rangle$. When the inner product is the standard dot product in $\mathbb{R}^n$, this adjoint operator is simply the matrix transpose, $\mathbf{J}^\mathrm{T}$.} \added{Its coefficient matrix is the transpose of the global Jacobian of the forward problem.} If the global Jacobian was explicitly assembled during the forward problem or if it is symmetric, the solver strategies mentioned in the first step can be safely reused. However, if the Jacobian is asymmetric and not explicitly assembled, solving the adjoint system iteratively requires the vector-Jacobian product (VJP) via backward-mode automatic differentiation of the \deleted{residual vector $f$} \added{residual vector with entries $f(\mathbf{u},v,\mathbf{N}_i)$} at each iteration. This process is generally far more computationally expensive than assembling an explicit sparse Jacobian and computing standard matrix-vector products. 
    \item (\deleted{The inverse problem }\added{Design sensitivity}) \deleted{Substituting the solutions $\mathbf{u}=\tilde{\mathbf{u}}$ and $\boldsymbol{\lambda}=\tilde{\boldsymbol{\lambda}}$ into Eq.~\eqref{eq:lagrange_solution}b yields the final gradient of the objective function with respect to the design variable $v$.} \added{Substituting the solutions $\mathbf{u}=\mathbf{u}_{\mathrm{D}}+\sum_j\tilde{U}_j\mathbf{N}_j$ and $\boldsymbol{\lambda}=\sum_i\tilde{\Lambda}_i\mathbf{N}_i$, and the basis functions $N^v_k$ of $\mathcal{D}_h$ as directions, into the left-hand side of Eq.~\eqref{eq:lagrange_solution}b yields the final gradient of the objective function with respect to the design variable, i.e., the column vector of nodal sensitivities $\nabla_{\mathbf{v}}\mathcal{L}$ with entries $D_v\mathcal{J}(\mathbf{u},v;N^v_k)+D_vf(\mathbf{u},v,\boldsymbol{\lambda};N^v_k)$, where $\mathbf{v}$ is the column vector of nodal design variables.} 
\end{enumerate}

The discretized design variable $\mathbf{v}$ is subsequently updated via gradient descent:
\begin{equation}
    \mathbf{v}\leftarrow \mathbf{v}-\eta \nabla_{\mathbf{v}}\mathcal{L}
\end{equation}
where $\eta$ is a step size (or learning rate) that can be dynamically scaled to accelerate convergence when stable, or reduced to prevent divergence. For optimization problems without inequality constraints, the Barzilai-Borwein method \cite{barzilai1988TwoPointStepSize} is employed. This approach approximates quasi-Newton methods by capturing the effect of the Hessian through a single scalar step size $\eta$ that is updated at each iteration. The update of $\eta$ incorporates exponential smoothing with a smoothing factor of 0.99, which represents the weight applied to the step size from the previous iteration. Alternative optimization algorithms, such as L-BFGS-B or Adam, can also be utilized for the parameter update; however, due to the non-convex nature of topology optimization, different optimizers may converge to distinct local minima.

When the optimization problem incorporates inequality constraints:
\begin{equation}
\begin{aligned}
\min_v\ &\mathcal{J}(\mathbf{u},v), \\
\mathrm{s.t.\ } &\text{\deleted{$\displaystyle f(\mathbf{u};v)=0,$}}\delcr\delamp\added{f(\mathbf{u},v,\delta\mathbf{u})=0\quad\forall\,\delta\mathbf{u}\in\mathcal{V}_h,} \\
& g_i(\mathbf{u},v) \leq 0, \quad i=1,2,\dots
\end{aligned}
\end{equation}
The sensitivity of $g_i$ with respect to $v$ can be efficiently computed by solving an analogous adjoint system for each constraint. Subsequently, the optimization process is driven using algorithms such as the Method of Moving Asymptotes (MMA) \cite{holmberg2013StressConstrainedTopology,svanberg1987MethodMovingAsymptotes,svanberg2002class}. MMA approximates the objective function $\mathcal{J}$ and constraint functions $g_i$ via locally convex surrogates, adapts their curvature based on the historical oscillation of the design variables, and computes the update by solving the Lagrangian dual problem. In this work, a variant of MMA enhanced by the Conservative Convex Separable Approximation (CCSA) \cite{svanberg2002class} is implemented following the implementation of the \texttt{NLopt} package \cite{NLopt}. This formulation evaluates the updated design variables to verify the conservativeness of the surrogate models, thereby ensuring stricter adherence to the inequality constraints. The update rule for the design variables is expressed as:
\begin{equation}
    \mathbf{v}\leftarrow \texttt{MMA}\left(\mathbf{v}, \mathcal{J}, g_i, \nabla_{\mathbf{v}}\mathcal{L}, \nabla_{\mathbf{v}} g_i\right).
\end{equation}

\subsection{Adjoint-method-based topology optimization}
\label{sec:simp}

In this work, the continuous composition ratio \added{$c$} is optimized either independently or concurrently with the structural topology \added{$\rho$}. While applying the adjoint state method to optimize the composition ratio field alone is relatively straightforward, topology optimization requires specific theoretical and algorithmic treatments. 

\subsubsection{Standard formulation}

This section introduces the classic Solid Isotropic Material with Penalization (SIMP) method \cite{bendsoe1989optimal} for topology optimization. This approach parameterizes the structural topology using a continuous density field $\rho$ (which can be optimized concurrently with the composition ratio $c$) to indicate material presence at any given spatial location. The objective function and constraints are written as
\begin{equation}
\begin{aligned}
\min_{\rho\added{,c}}&\ \mathcal{J}(\mathbf{u},\rho\added{,c}) = 
\mathfrak{N} \underset{\mathrm{Stored\ energy}}{\underbrace{\left[\int_\Omega \rho^\varrho \psi(\mathbf{u}\added{,c})\mathrm{d}V\right]}},\\
\mathrm{s.t.}&\delsp\text{\deleted{$\displaystyle f(\mathbf{u};\rho)=D_\mathbf{u} \left(\int_\Omega \rho^\varrho\psi(\mathbf{u})\mathrm{d}V-\int_{\partial \Omega} \mathbf{t}\cdot \mathbf{u}\mathrm{d}A\right)=0,$}}\delcr
\delamp\ \added{f(\mathbf{u},\rho,c,\delta\mathbf{u})=D_\mathbf{u}\Pi(\mathbf{u},\rho,c;\delta\mathbf{u})}\\
&\ \added{\phantom{f(\mathbf{u},\rho,c,\delta\mathbf{u})}=\int_\Omega \rho^\varrho\,\mathbf{P}(\mathbf{u},c):\nabla\delta\mathbf{u}\,\mathrm{d}V-\int_{\partial \Omega} \mathbf{t}\cdot \delta\mathbf{u}\,\mathrm{d}A=0\quad\forall\,\delta\mathbf{u}\in\mathcal{V}_h,}\\
&\ \added{\text{with}\ \Pi(\mathbf{u},\rho,c)=\int_\Omega \rho^\varrho\psi(\mathbf{u},c)\mathrm{d}V-\int_{\partial \Omega} \mathbf{t}\cdot \mathbf{u}\mathrm{d}A,}\\
&\ g(\rho\added{,c})=\int_{\Omega} \left(\rho -\theta\right) \mathrm{d}V \leq 0,
\label{eq:simp_objective}
\end{aligned}
\end{equation}
where $\rho\in[\rho_{\min}, 1]$ is the bounded density field with a small positive lower limit to prevent stiffness matrix singularities, \added{$c\in[0,1]$ is the composition ratio, bounded to the range where the pICNN is trained and can interpolate well, }$\theta<1$ is the target volume fraction, and $\mathfrak{N}[x]$ is an operator that normalizes the enclosed value $x$ to its initial value before the first iteration $x_{\mathrm{init}}$ ($\mathfrak{N}[x]=x/x_{\mathrm{init}}$). Furthermore, $\psi$ represents the strain energy density, which can originate from a standard phenomenological model or be derived from experimental data via machine learning approaches, as formulated in Eq.~\eqref{eq:total_strain_energy_NN_soft}. \added{The equilibrium constraint is the weak form of the state equation in Eq.~\eqref{eq:constrained_min}, obtained as the Gâteaux derivative of the total potential energy $\Pi$ in the direction of the test function $\delta\mathbf{u}\in\mathcal{V}_h$, where $\mathbf{P}=\partial\psi/\partial\mathbf{F}$ is the first Piola-Kirchhoff stress, $\nabla$ denotes the gradient with respect to the reference coordinates, and $\mathbf{t}$ is the prescribed traction.}

Following the standard SIMP approach \cite{bendsoe1989optimal,sigmund200199}, the strain energy density is scaled by the density field via $\rho^\varrho \psi$. Here, $\varrho \in [2, 5]$ serves as the penalization factor, which makes intermediate density values mechanically inefficient relative to their volume contribution, thereby driving the optimized density field toward the discrete binary bounds of 0 and 1. To ensure stable convergence, a continuation method is employed: once the optimization converges for a given $\varrho$, the penalty factor is incrementally increased based on the fraction of remaining intermediate 'gray' material until the predefined upper bound is reached (as detailed in Section \ref{sec:implementation}).

\subsubsection{Special treatment for finite deformation}

To enhance numerical stability under finite deformation, the following strategies are employed, specifically tailored to mitigate numerical artifacts within finite element frameworks like \texttt{FEniCSx} that rely on automated symbolic differentiation. 

\textbf{Material interpolation in void regions}. As the density or stiffness of certain regions decreases during topology optimization, elements within these void regions may become severely distorted and ultimately cause non-convergence. This issue is particularly pronounced in our near-incompressible formulation, where the stiff volumetric penalty term is highly sensitive to element distortion. To alleviate this, a widely used method \cite{wang2014InterpolationSchemeFictitious} interpolates between the original material model and a linear elastic model:
\begin{equation}
\begin{aligned}
    &\text{\deleted{$\displaystyle \psi_{\mathrm{interp}}(\mathbf{u},c,\mathcal{I})=\psi(\mathcal{I}\mathbf{u},c)-\psi_{L}(\mathcal{I}\mathbf{u})+\psi_L (\mathbf{u}),$}}\delcr
    \delamp\added{\psi_{\mathrm{interp}}(\mathbf{u},c,\mathcal{I})=\psi(\mathcal{I}\mathbf{u},c)-\psi_{L}(\mathcal{I}\mathbf{u},c)+\psi_L (\mathbf{u},c),}
\end{aligned}
\end{equation}
where $\psi_{L}=\frac{1}{2}\lambda(\boldsymbol{\varepsilon}_{kk})^2+\mu\boldsymbol{\varepsilon}_{ij}\boldsymbol{\varepsilon}_{ij}$ is the strain energy density for linear elasticity (with Lamé constants $\lambda$ and $\mu$\added{ computed from $E(c)$ in Eq.~\eqref{eq:fitted-E-c} and $\nu$}, and the infinitesimal strain tensor $\boldsymbol{\varepsilon}$), and $\mathcal{I}$ is a smoothed indicator field identifying the void regions:
\begin{equation}
    \mathcal{I}=\frac{\tanh{(\theta_{\mathrm{thres}}\theta_{\mathrm{sharp}})}+\tanh{(\theta_{\mathrm{sharp}}(\rho^{\varrho}-\theta_{\mathrm{thres}}))}}{\tanh{(\theta_{\mathrm{thres}}\theta_{\mathrm{sharp}})}+\tanh{(\theta_{\mathrm{sharp}}(1-\theta_{\mathrm{thres}}))}},
    \label{eq:void-indicator}
\end{equation}
where $\theta_{\mathrm{thres}}=0.01$ and $\theta_{\mathrm{sharp}}=500$, consistent with the literature \cite{wang2014InterpolationSchemeFictitious}. While this expression is functionally equivalent to a standard sigmoid, this specific formulation prevents numerical overflow by utilizing $\tanh(\cdot)$ rather than explicit $\exp(\cdot)$ evaluations. As demonstrated in previous studies \cite{wang2014InterpolationSchemeFictitious,deng2025DatadrivenMultiscaleNonlinear}, this linear interpolation scheme proves effective for cantilever and double-clamped beams. In such scenarios, void regions undergoing large deformations are not fully constrained by boundary conditions or surrounding solid material; consequently, element overlapping and inversion are numerically tolerated, as the elements can deform freely without imparting significant fictitious stresses back into the primary structure.

However, when modeling near-incompressible behaviors where void regions are fully encapsulated by solid material, this linear elastic interpolation becomes fundamentally flawed. Because the infinitesimal strain tensor is not rotationally invariant, it erroneously generates artificial stresses during finite rigid-body rotations. This renders the linear elastic model kinematically incompatible with the primary hyperelastic material, frequently triggering solver divergence once void regions coalesce and the interpolation actively engages. Even if a Saint Venant-Kirchhoff model (which utilizes the rotationally invariant Green-Lagrange strain tensor) is adopted, severe element inversion instabilities under compression could occur, similarly resulting in non-convergence. Therefore, a simple neo-Hookean model is utilized as the robust base for interpolation:
\begin{equation}
    \psi_L(\mathbf{F}\added{,c}) = \frac{1}{2}\mu\left(J^{-2/3}I_1-3\right)+\frac{1}{2}\kappa\left(J-1\right)^2.
\end{equation}
Note that this formulation omits natural logarithm terms, ensuring the energy remains well-defined across all possible strain invariants, even when elements become severely distorted. 

Furthermore, computing the strain directly from the penalized displacement field $\mathcal{I}\mathbf{u}$ introduces severe numerical instability. Because both $\mathcal{I}$ and $\mathbf{u}$ depend on the spatial coordinates, the spatial gradient of $\mathcal{I}$ generates massive spurious strain components due to the steepness of the indicator function. In in-house finite element implementations, this issue is typically circumvented by explicitly omitting the $\mathbf{u} \otimes \nabla \mathcal{I}$ term from the tangent stiffness derivation. However, within frameworks like \texttt{FEniCSx} that rely on exact automated symbolic differentiation, it is computationally sufficient to directly scale the displacement gradient instead:
\begin{equation}
\begin{aligned}
    &\text{\deleted{$\displaystyle \psi_{\mathrm{interp}}(\mathbf{F},c,\mathcal{I})= \psi\left(\mathbf{I}+\mathcal{I}\frac{\partial \mathbf{u}}{\partial \mathbf{X}},c\right)-\psi_L\left(\mathbf{I}+\mathcal{I}\frac{\partial \mathbf{u}}{\partial \mathbf{X}}\right)+\psi_L\left(\mathbf{F}\right).$}}\delcr
    \delamp\added{\psi_{\mathrm{interp}}(\mathbf{F},c,\mathcal{I})= \psi\left(\mathbf{I}+\mathcal{I}\frac{\partial \mathbf{u}}{\partial \mathbf{X}},c\right)-\psi_L\left(\mathbf{I}+\mathcal{I}\frac{\partial \mathbf{u}}{\partial \mathbf{X}},c\right)+\psi_L\left(\mathbf{F},c\right).}
\end{aligned}
\end{equation}

In addition, this fictitious material interpolation unintentionally interferes with the optimization landscape. Although the energy interpolation is strictly a numerical regularization technique to preserve solver robustness and should not dictate the optimal structural design, it inevitably alters the structural sensitivities via the adjoint state method. This introduces a severe numerical artifact: because the void indicator $\mathcal{I}$ depends on the density $\rho$, the optimizer can exploit this dependency to bias the structural solution toward the fictitious interpolated model simply because it artificially yields a lower objective value. Consequently, the void indicator must be explicitly decoupled from the differentiation chain rule, meaning the adjoint solver must be prohibited from evaluating the sensitivity of $\mathcal{I}$ with respect to $\rho$. This decoupling is achieved by applying an operation to $\mathcal{I}$ that effectively blocks the symbolic differentiation graph without altering its numerical value—specifically, by projecting $\mathcal{I}$ onto the finite element function space via an independent linear solve prior to the objective evaluation. 

\textbf{Lower bound of material penalization in topology optimization}. In scenarios involving compression, buckling can occur within the optimized members. The forces triggering buckling often originate from spurious artificial stresses within the void regions or incremental density updates. If artificial buckling occurs, the structure may uncontrollably oscillate between different bifurcation branches during each density update, causing severe convergence failures for the nonlinear solver and substantially obstructing convergence. To alleviate this, the applied boundary loading is restricted to a moderate magnitude to prevent such instances, while simultaneously enforcing a rigid lower bound of $0.001$ on the stiffness penalization term $\rho^{\varrho}$ to suppress these void-induced numerical perturbations.

\subsection{Filtering of topology and composition ratio distribution}
\label{sec:filter}

In topology optimization, the optimized density field may exhibit the \textit{checkerboard effect}, a numerical artifact where the finite element solution converges to an artificially stiff, alternating pattern of solid and void elements that is physically unrealistic and indicative of spurious numerical oscillations \cite{lazarov2011FiltersTopologyOptimization}. Furthermore, the solutions often suffer from mesh dependence: refining the finite element mesh yields a fundamentally different structural topology rather than the same design simply at a higher resolution. Similar numerical instabilities can manifest in other continuous design variables, such as the composition ratio field investigated in this work.  

To address these issues, one of three strategies is typically employed: modifying the gradient or density (the one-field approach), blurring the design field to create a new physical field for material penalization and volume computation (the two-field approach), or further projecting the blurred field using a sharpening function to yield a third physical field (the three-field approach) \cite{sigmund2013topology}. Formally, the three-field approach treats a variable $v$ as an underlying numerical design field, applies a spatial filter to obtain a blurred field $\tilde{v}$, and projects $\tilde{v}$ onto a sharper physical field $\hat{v}$. This underlying field $v$ can represent the density field \added{$\rho$} for topology optimization, other continuous design variables (such as the composition ratio \added{$c$}) \cite{wei2026MachineLearningbasedMultiscale,vijayakumaran2025ConsistentMachineLearninga}, or both simultaneously.  

This work utilizes a PDE-based filtering approach \cite{lazarov2011FiltersTopologyOptimization}, which formulates the blurring operation as the solution to a Helmholtz-type differential equation. The blurred field $\tilde{v}$ and the underlying field $v$ satisfy:
\begin{equation}
\begin{aligned}
-r_{\mathrm{f}}^2\nabla^2 \tilde{v} + \tilde{v}&=v\quad\text{in}\ \Omega,\\
\nabla\tilde{v}\cdot \mathbf{n}&=0\quad \text{on}\ \partial\Omega,
\end{aligned} 
\end{equation}
where $r_{\mathrm{f}}$ is the filter length scale parameter, and $\mathbf{n}$ is the outward surface normal vector. This PDE is expressed in its weak form\added{, with $\tilde{v}\in\mathcal{D}_h$ and test functions $\delta v\in\mathcal{D}_h$,} as:
\begin{equation}
    \int_{\Omega}\left(r_{\mathrm{f}}^2\nabla\tilde{v}\cdot\nabla\delta v+\tilde{v}\delta v-v\delta v \right)\mathrm{d}V - \int_{\partial\Omega}r_{\mathrm{f}}^2(\nabla\tilde{v}\cdot \mathbf{n})\delta v\mathrm{d}A=0\text{\deleted{.}}\added{\quad\forall\,\delta v\in\mathcal{D}_h.}
\end{equation}
Applying the Neumann boundary condition causes the boundary integral to vanish, yielding:
\begin{equation}
    \int_{\Omega}\left(r_{\mathrm{f}}^2\nabla\tilde{v}\cdot\nabla\delta v+\tilde{v}\delta v-v\delta v \right)\mathrm{d}V=0\text{\deleted{.}}\added{\quad\forall\,\delta v\in\mathcal{D}_h.}
\end{equation}
Consequently, the global coefficient matrix only needs to be assembled once at the beginning of the topology optimization process. The discretized linear system is expressed as:
\begin{equation}
    \underset{\mathbf{K}_{\mathrm{f}}}{\underbrace{\bigcup \left[ \int_{\Omega_e}\left(r_{\mathrm{f}}^2\nabla\mathbf{N}_e^\mathrm{T}\nabla\mathbf{N}_e+\mathbf{N}_e^{\mathrm{T}}\mathbf{N}_e\right)\mathrm{d}V \right]}} \tilde{\mathbf{v}} = \underset{\mathbf{M}_{\mathrm{f}}}{\underbrace{\bigcup \left[ \int_{\Omega_e} \mathbf{N}_e^{\mathrm{T}}\mathbf{N}_e\mathrm{d}V \right]}} \mathbf{v},
\end{equation}
where $\bigcup$ denotes the global finite element assembly operator, $\mathbf{N}_e$ represents the element shape function matrix, and $\tilde{\mathbf{v}}$ and $\mathbf{v}$ are the global nodal vectors of $\tilde{v}$ and $v$, respectively. The assembled global coefficient matrices are thus defined as the filter stiffness matrix $\mathbf{K}_{\mathrm{f}}$ and the mass matrix $\mathbf{M}_{\mathrm{f}}$. 

When evaluating the total gradient $\nabla_\mathbf{v} \mathcal{L}$, the combination of the adjoint method and symbolic differentiation allows us to directly compute the exact gradient with respect to the filtered field, $\nabla_{\tilde{\mathbf{v}}} \mathcal{L}$. However, this leaves the final step of the chain rule, $\partial \tilde{\mathbf{v}}/\partial \mathbf{v}$, which is governed by the linear system solved during the forward filtering pass:
\begin{equation}
\frac{\partial \tilde{\mathbf{v}}}{\partial \mathbf{v}}=\mathbf{K}_{\mathrm{f}}^{-1}\mathbf{M}_{\mathrm{f}}.
\end{equation}
Exploiting the symmetry of both $\mathbf{K}_{\mathrm{f}}$ and $\mathbf{M}_{\mathrm{f}}$, the final gradient is formulated as:
\begin{equation}
    \nabla_{\mathbf{v}} \mathcal{L}=\left(\mathbf{K}_{\mathrm{f}}^{-1}\mathbf{M}_{\mathrm{f}}\right)^\mathrm{T}\nabla_{\tilde{\mathbf{v}}} \mathcal{L}=\mathbf{M}_{\mathrm{f}}^\mathrm{T}\mathbf{K}_{\mathrm{f}}^{-\mathrm{T}}\nabla_{\tilde{\mathbf{v}}} \mathcal{L}=\mathbf{M}_{\mathrm{f}}\mathbf{K}_{\mathrm{f}}^{-1}\nabla_{\tilde{\mathbf{v}}} \mathcal{L}.
\end{equation}
\deleted{In practice, this is efficiently evaluated by first solving the adjoint linear system $\mathbf{K}_{\mathrm{f}}\boldsymbol{\lambda}_{\mathrm{f}}=\nabla_{\tilde{\mathbf{v}}} \mathcal{L}$ for the intermediate adjoint vector $\boldsymbol{\lambda}_{\mathrm{f}}$, and subsequently applying the mass matrix projection $\nabla_{\mathbf{v}} \mathcal{L}=\mathbf{M}_{\mathrm{f}}\boldsymbol{\lambda}_{\mathrm{f}}$.} \added{In practice, this is efficiently evaluated by first solving the adjoint linear system $\mathbf{K}_{\mathrm{f}}\boldsymbol{\Lambda}_{\mathrm{f}}=\nabla_{\tilde{\mathbf{v}}} \mathcal{L}$ for the intermediate adjoint vector $\boldsymbol{\Lambda}_{\mathrm{f}}$, and subsequently applying the mass matrix projection $\nabla_{\mathbf{v}} \mathcal{L}=\mathbf{M}_{\mathrm{f}}\boldsymbol{\Lambda}_{\mathrm{f}}$.}

When the underlying design variable is the density field $\rho$, the three-field approach is employed. To promote a sharp structural boundary suitable for manufacturing, the blurred density $\tilde{\rho}$ is projected onto a physical density field $\hat{\rho}$ using a smooth Heaviside function:
\begin{equation}
    \hat{\rho}=\frac{\tanh{(\beta\tau)}+\tanh{(\beta(\tilde{\rho}-\tau))}}{\tanh{(\beta\tau)}+\tanh{(\beta(1-\tau))}},
    \label{eq:density-projection}
\end{equation}
where $\beta=2^{\varrho}$ controls the projection sharpness, and the projection threshold is set to $\tau=0.5$. Consequently, the optimization problem is reformulated using the projected physical density $\hat{\rho}$:
\begin{equation}
\begin{aligned}
\min_{\rho\added{,c}}&\ \mathcal{J}(\mathbf{u},\rho\added{,c}) = 
\mathfrak{N} \underset{\mathrm{Stored\ energy}}{\underbrace{\left[\int_\Omega \hat{\rho}^\varrho \psi(\mathbf{u}\added{,c})\mathrm{d}V\right]}},\\
\mathrm{s.t.}&\delsp\text{\deleted{$\displaystyle f(\mathbf{u};\rho)=D_\mathbf{u} \left(\int_\Omega \hat{\rho}^\varrho\psi(\mathbf{u})\mathrm{d}V-\int_{\partial \Omega} \mathbf{t}\cdot \mathbf{u}\mathrm{d}A\right)=0,$}}\delcr
\delamp\ \added{f(\mathbf{u},\rho,c,\delta\mathbf{u})=D_\mathbf{u}\Pi(\mathbf{u},\rho,c;\delta\mathbf{u})=0\quad\forall\,\delta\mathbf{u}\in\mathcal{V}_h,}\\
&\ \added{\text{with}\ \Pi(\mathbf{u},\rho,c)=\int_\Omega \hat{\rho}^\varrho\psi(\mathbf{u},c)\mathrm{d}V-\int_{\partial \Omega} \mathbf{t}\cdot \mathbf{u}\mathrm{d}A,}\\
&\ g(\rho\added{,c})=\int_{\Omega} \left(\hat{\rho} -\theta\right) \mathrm{d}V \leq 0.
\end{aligned}
\end{equation}
Similarly, the projected density $\hat{\rho}$ replaces $\rho$ in the void indicator defined in Eq.~\eqref{eq:void-indicator}. 

This PDE-based filtering is concurrently applied to the composition ratio $c$ following the two-field approach. \added{Accordingly, the filtered field $\tilde{c}$ replaces $c$ wherever the composition ratio enters $\psi$ and the constraints, so that $f$, $\mathcal{J}$, and $g_i$ depend on the design fields $\rho$ and $c$ through $\hat{\rho}$ and $\tilde{c}$.} This smooths the spatial transition between different constituent materials, thereby enhancing material compatibility and overall structural integrity \cite{pragya2023SoftFunctionallyGradient}.

\subsection{Implementation using symbolic-differentiation-based \texttt{FEniCSx}}
\label{sec:implementation}

When incorporating machine-learning-based constitutive laws alongside complex optimization objectives and constraints, deriving the analytical expressions required for both the forward problem (solving the governing PDEs) and the \deleted{inverse problem} \added{design sensitivity analysis} (evaluating adjoint-based gradients) becomes highly laborious, particularly during the rapid prototyping of new formulations. In this work, the finite element analysis and adjoint state method are implemented using \texttt{FEniCSx} (version 0.9) \cite{Baratta_DOLFINx_the_next_2023}. This framework relies on automated symbolic differentiation, requiring only the weak form of the PDEs and the explicit definitions of the objective and constraint functions. These are formulated using the Unified Form Language (UFL), which natively supports the exact symbolic differentiation of all built-in operators.

Nevertheless, situations arise where automated symbolic differentiation cannot traverse implicit operations within the differentiation chain, such as computing the sensitivity of the blurred field $\tilde{v}$ with respect to the underlying design field $v$. In such cases, the analytical sensitivities are manually derived and implemented to supplement the symbolically generated gradients. \added{Apart from this filter chain rule and the void indicator $\mathcal{I}$, which is held fixed in the differentiation (Section~\ref{sec:simp}), all derivatives in this work, including those of the constitutive law, the objectives, and the constraints, are generated symbolically. Their explicit expressions are listed in \ref{sec:appendix-sensitivity} for reference.}

The coupled optimization problem for the structural topology and material composition distribution is solved according to Algorithm \ref{algo:simp}. \added{The optimization is restricted to $[\rho_{\min},1]$ for $\rho$ where it is defined, and $[0,1]$ for $c$ where it is normalized and the pICNN is trained and can interpolate accurately ($R^2\geq0.96$ for the held-out composition $c=0.1755$).} If either the topology or the composition ratio optimization is omitted, the algorithmic steps related to updating the corresponding design variable are bypassed. While all numerical examples are demonstrated in 2D under the plane strain assumption to reduce computational cost, the theoretical formulations and UFL-based implementations are intrinsically dimension-independent and directly applicable to 3D problems without modification. 

\begin{algorithm}[!htbp]
\caption{Optimization loop for structural topology and material composition}
\label{algo:simp}
\small
\setstretch{1}
\SetAlgoLined

Initialize \deleted{underlying} design fields $\rho=1.0$, $c=0.5$\added{ (unless otherwise stated)}, and penalty factor $\varrho=2$\;
\For{each iteration}{
Apply the PDE filter by solving the Helmholtz equations for $\tilde{\rho}$ and $\tilde{c}$\;
Project $\tilde{\rho}$ onto the physical density $\hat{\rho}$ using the smooth Heaviside function\;
\deleted{Solve the nonlinear hyperelastic equilibrium $f(\mathbf{u};\hat{\rho},\tilde{c})=0$ for the state variable $\mathbf{u}$} \added{Solve the nonlinear hyperelastic equilibrium $f(\mathbf{u},\rho,c,\delta\mathbf{u})=0$, $\forall\,\delta\mathbf{u}\in\mathcal{V}_h$ for the nodal displacement vector $\mathbf{U}$}\;
\deleted{Compute the objective $\mathcal{J}(\mathbf{u},\hat{\rho},\tilde{c})$ and constraints $g_i(\hat{\rho}, \tilde{c})$} \added{Compute the objective $\mathcal{J}(\mathbf{u},\rho,c)$ and constraints $g_i(\mathbf{u},\rho,c)$}\;
\deleted{Solve the adjoint equations $\left(\nabla_\mathbf{u} f\right)^\mathrm{T}\boldsymbol{\lambda}_{(\cdot)} = -\nabla_\mathbf{u}(\cdot)$ to obtain the adjoint variables, where $(\cdot)$ represents the objective $\mathcal{J}$ or inequality constraints $g_i$} \added{Solve the adjoint equations $\mathbf{J}^\mathrm{T}\boldsymbol{\Lambda}_{(\cdot)} = -\nabla_\mathbf{U}(\cdot)$ to obtain the nodal vectors of the adjoint variables, where $(\cdot)$ represents the objective $\mathcal{J}$ or inequality constraints $g_i$}\;
\deleted{Compute the objective sensitivities $\nabla_\rho \mathcal{L}$, $\nabla_c \mathcal{L}$, and constraint sensitivities $\nabla_\rho g_i$, $\nabla_c g_i$ via the adjoint state method and the filter chain rule} \added{Compute the objective sensitivities $\nabla_{\boldsymbol{\rho}} \mathcal{L}$, $\nabla_{\mathbf{c}} \mathcal{L}$, and constraint sensitivities $\nabla_{\boldsymbol{\rho}} g_i$, $\nabla_{\mathbf{c}} g_i$ with respect to the nodal vectors $\boldsymbol{\rho}$ and $\mathbf{c}$ of $\rho$ and $c$ via the adjoint state method and the filter chain rule}\;
Restrict all sensitivities to the range $[-0.01, 0.01]$\;
\deleted{Update $\rho$ using MMA: $\rho\leftarrow \texttt{MMA}\left (\rho, \mathcal{J}, g_i, \nabla_\rho\mathcal{L}, \nabla_\rho g_i\right)$} \added{Update $\boldsymbol{\rho}$ using MMA: $\boldsymbol{\rho}\leftarrow \texttt{MMA}\left (\boldsymbol{\rho}, \mathcal{J}, g_i, \nabla_{\boldsymbol{\rho}}\mathcal{L}, \nabla_{\boldsymbol{\rho}} g_i\right)$}\;
\deleted{Update $c$ using MMA: $c\leftarrow \texttt{MMA}\left (c, \mathcal{J}, g_i, \nabla_c\mathcal{L}, \nabla_c g_i\right)$, or via gradient descent with a Barzilai-Borwein step size $\eta$: $c\leftarrow c-\eta \nabla_c\mathcal{L}$} \added{Update $\mathbf{c}$ using MMA: $\mathbf{c}\leftarrow \texttt{MMA}\left (\mathbf{c}, \mathcal{J}, g_i, \nabla_{\mathbf{c}}\mathcal{L}, \nabla_{\mathbf{c}} g_i\right)$, or via gradient descent with a Barzilai-Borwein step size $\eta$: $\mathbf{c}\leftarrow \mathbf{c}-\eta \nabla_{\mathbf{c}}\mathcal{L}$}\;
\deleted{Enforce constraints $\rho\in[\rho_{\min}, 1]$ and $c\in [0,1]$} \added{Enforce the bounds $\rho_{\min}\leq\rho_k\leq1$ and $0\leq c_k\leq1$ on all entries of $\boldsymbol{\rho}$ and $\mathbf{c}$}\;
\If{the relative objective value decreases by less than \texttt{TOL} in two consecutive iterations or has not decreased for 50 iterations after 500 iterations}{
\eIf{$\varrho<\varrho_{\max}$ and topology optimization is activated}{
Compute the intermediate density fraction $\gamma=\frac{4}{\texttt{Vol}}\int_\Omega \hat{\rho}(1-\hat{\rho})\mathrm{d}V$\added{, with $\texttt{Vol}=\int_\Omega\mathrm{d}V$}\;
Update the penalty factor: $\varrho\leftarrow \min\left\{\varrho\left(1+0.2^{1+\gamma/2}\right),\varrho_{\max}\right\}$\;
}{
Terminate the optimization\;
}
}
}
\end{algorithm}

\section{Results}
\label{sec:results}

In this section, the proposed computational design workflow is applied to problems relevant to 3D printing multimaterial components for robotic grippers for object manipulation, a domain that traditionally relies heavily on prototyping. The focus is on two classes of problems: (1) Design contact points, such as fingertips, for robotic hands with tunable anisotropic responses \added{utilizing the distinct stiffnesses of material components}, and (2) lightweight resilient design of fingers for robotic grippers that accommodate large deformations, \deleted{taking into account material limitations} \added{utilizing the exceptional elongation tolerance of one of the material components (Agilus)}.  
The contact point problem is first analyzed within the infinitesimal strain regime, where only the continuous composition ratio field is optimized to achieve a target anisotropic response (Section \ref{sec:contact-point-small-deform}). This framework is subsequently extended to the finite deformation regime under more general contact conditions utilizing third medium contact theory (Section \ref{sec:contact-points-large-deform}). The second example for the fingers of the robotic gripper addresses the concurrent optimization of macroscopic topology and digital material distribution (Section \ref{sec:clamping-fingers}), subject to an ultimate tensile stretch constraint, to demonstrate the distinct efficacy and advantages of this \deleted{multiscale}\added{concurrent} optimization approach. Specifically, four different optimization scenarios are considered for this multi-objective optimization problem.

\subsection{Contact points: Composition optimization for anisotropic response under small deformation}
\label{sec:contact-point-small-deform}

\begin{figure}[ht]
    \centering
    \includegraphics[width=0.9\linewidth]{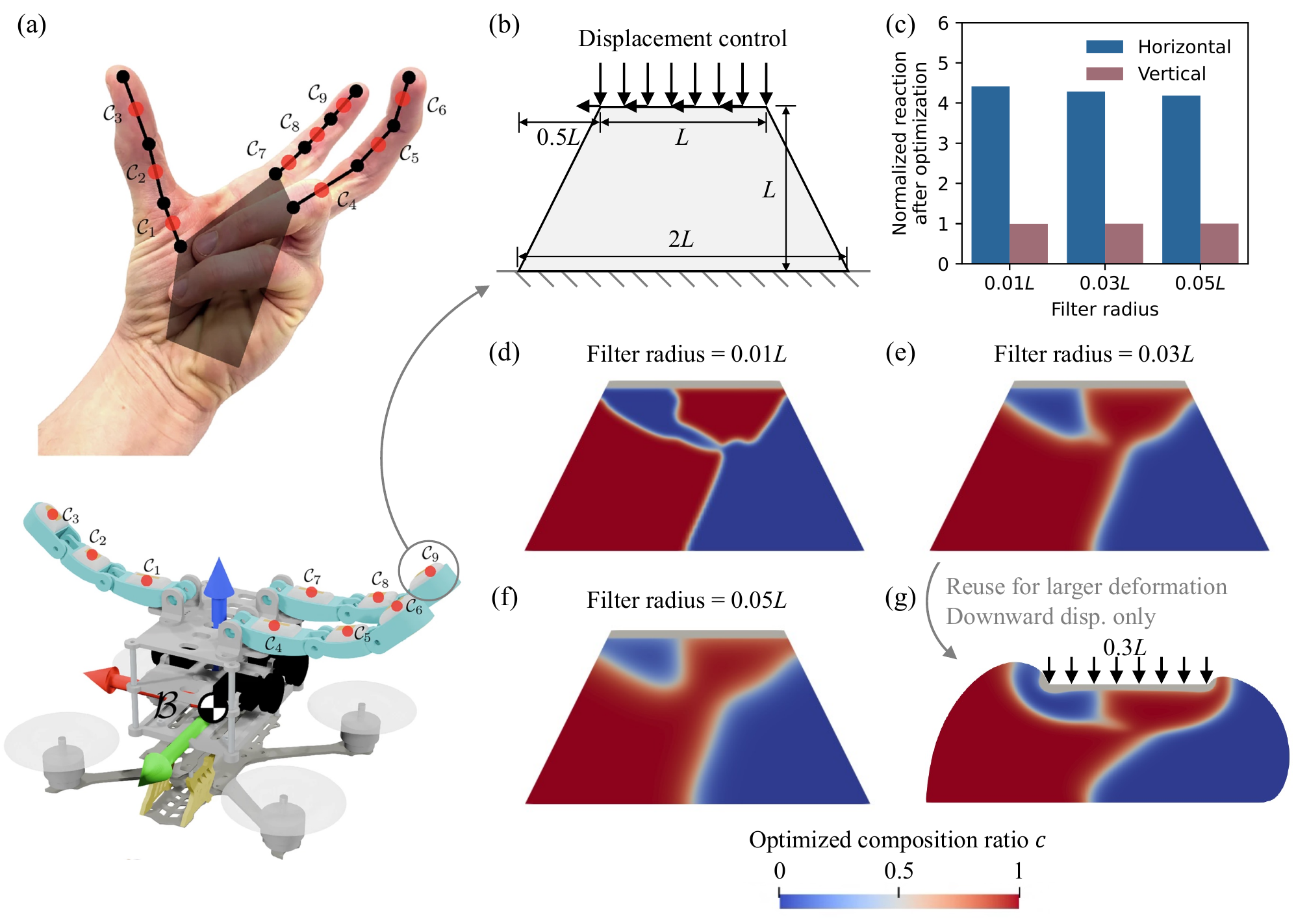}
    \caption{Optimization of the continuous composition ratio $c$ within a trapezoidal domain, modeling a localized contact point for a robotic gripper, to achieve an anisotropic mechanical response. \added{(a) A human hand (top) grasps objects of varying shapes while providing tactile feedback, whereas a robotic gripper (bottom) serves this purpose with the same antagonistic configuration and nine soft pads as contact points. Images adapted from ref.~\cite{bredenbeck2026aerial} under a CC BY 4.0 license. }\deleted{(a)}\added{(b)} Geometry and boundary conditions of the trapezoidal domain\added{, mimicking one of the soft pads}. \deleted{(b)}\added{(c)} Horizontal and vertical reaction forces on the top boundary after optimization, normalized with respect to the initial homogeneous design. \deleted{(c)-(e)}\added{(d)-(f)} Optimized distributions of the composition ratio $c$ utilizing various filter radii to control the spatial transition gradient between material phases. \deleted{(f)}\added{(g)} Structural response of the optimized design evaluated under finite deformation, subjected to a prescribed downward displacement. }
    \label{fig:trapezoid_optimize_with_threshold}
\end{figure}

The first numerical example, a two-dimensional trapezoidal domain shown in Fig.~\ref{fig:trapezoid_optimize_with_threshold}\deleted{(a)}\added{(b)}, models a localized \added{soft }contact point on a \added{stiff }robotic gripper \added{that mimics a human hand (Fig.~\ref{fig:trapezoid_optimize_with_threshold}(a)) \cite{bredenbeck2026aerial}}\deleted{\cite{hammond2012towards}}. To optimize grasping\added{ and sensing} performance, the contact point must exhibit a highly anisotropic mechanical response. Specifically, it requires high compliance in the normal direction to conformally adapt to the geometry of the target object under compression\added{ and to be sensitive to compressive loading as a force sensor}. Conversely, it must maintain high stiffness in the tangential direction to maximize shear resistance and securely transmit frictional forces. 

The geometry of the trapezoidal domain features a fixed bottom boundary measuring $2L$ in length, with an upper boundary length and total height both measuring $L$. The domain is discretized into a structured mesh comprising 9,408 linear quadrilateral plane-strain elements. The upper boundary is simultaneously subjected to a downward displacement of $0.01L$ and a leftward displacement of $0.001L$. The continuous composition ratio distribution  $c$ is optimized within this small deformation regime. It is noted that topology is not optimized in this example\added{ ($\rho\equiv1$)}, thus the initial geometry remains fixed and the material distribution $c$ is optimized. %Subsequently, the optimized design is evaluated under finite deformation loading scenarios. 
To prevent singularities near the prescribed displacement, the underlying design variable $c$ is constrained to 0.5 within a $0.04L$-thick layer immediately adjacent to the top boundary (although the filtered field $\tilde{c}$ remains free to vary). In the remaining design domain, the variable evolves from a uniform initial value of 0.5. The optimization problem is formulated as
\begin{equation}
\begin{aligned}
\min_c\ \mathcal{J}(\mathbf{u}\added{,\rho\equiv1},c) =& 
\mathfrak{N} \underset{\mathrm{Vertical\ reaction}}{\underbrace{\left[\left |\int_{\partial\Omega_{\text{top}}} \mathbf{P}\cdot \mathbf{n}\cdot \mathbf{e}_y \mathrm{d}A\right|\right]}} 
- \mathfrak{N} \underset{\mathrm{Horizontal\ reaction}}{\underbrace{ \left[\left|\int_{\partial\Omega_{\text{top}}} \mathbf{P}\cdot \mathbf{n}\cdot \mathbf{e}_x\mathrm{d}A\right|\right]}}+10\\
\text{\deleted{$\displaystyle \mathrm{s.t.}\ f(\mathbf{u};c)=$}}\delamp\text{\deleted{$\displaystyle D_{\mathbf{u}} \left(\int_\Omega\psi(\mathbf{u},c)\mathrm{d}V-\int_{\partial \Omega} \mathbf{t}\cdot \mathbf{u}\mathrm{d}A\right)=0,$}}\delcr
\added{\mathrm{s.t.}\ f(\mathbf{u},\rho\equiv1,c,\delta\mathbf{u})={}}&\added{D_{\mathbf{u}}\Pi(\mathbf{u},\rho\equiv1,c;\delta\mathbf{u})=0\quad\forall\,\delta\mathbf{u}\in\mathcal{V}_h,}\\
\added{\text{with}\ \Pi(\mathbf{u},\rho\equiv1,c)={}}&\added{\int_\Omega\psi(\mathbf{u},c)\mathrm{d}V-\int_{\partial \Omega} \mathbf{t}\cdot \mathbf{u}\mathrm{d}A,}
\label{eq:trapezoid_objective}
\end{aligned}
\end{equation}
where $\mathbf{P}$ is the first Piola-Kirchhoff stress tensor, $\mathbf{n}$ is the reference outward surface normal, and $\mathbf{e}_x$ and $\mathbf{e}_y$ are the Cartesian unit basis vectors. The addition of 10 to the objective function is to keep its value positive for the convergence criterion, with no effect on the optimization trajectory. 
Note that while the objective function is structured to simultaneously minimize the vertical reaction and maximize the horizontal reaction, this unconstrained scalarization primarily maximizes the relative difference between the two terms. Consequently, it does not strictly guarantee an absolute decrease in the vertical reaction or an absolute increase in the horizontal reaction independently. The initial step size for the Barzilai-Borwein update is set to 1, and the relative tolerance \texttt{TOL} (in Algorithm \ref{algo:simp}) for convergence criteria is set to 0.01\%. %Structural topology optimization is deactivated for this specific example.  

Three distinct filter radii ($0.01L$, $0.03L$, and $0.05L$) are employed to smooth the composition ratio field $c$, with the resulting optimized material distributions and reaction forces visualized in Fig.~\ref{fig:trapezoid_optimize_with_threshold}\deleted{(b)-(e)}\added{(c)-(f)}. After optimization, the horizontal reaction force increases to more than four times its initial magnitude, while the vertical reaction remains relatively constant, successfully fulfilling the desired design criteria. This highly anisotropic behavior is driven by the tailored material distribution (Fig.~\ref{fig:trapezoid_optimize_with_threshold}\deleted{(c)-(e)}\added{(d)-(f)}). The underlying kinematic mechanism becomes apparent when this optimized design is evaluated under finite deformation subjected purely to vertical compression, as shown in Fig.~\ref{fig:trapezoid_optimize_with_threshold}\deleted{(f)}\added{(g)}. Under pure vertical loading, the contact point exhibits a coupled lateral shear response. Two stiffer domains undergo localized shear deformation relative to one another, effectively forming a contact-aided compliant mechanism \cite{mankame2002contact} that provides enhanced resistance to horizontal forces. Furthermore, as the filter radius increases, the spatial transition within the $c$ field becomes smoother without significantly compromising the amplified horizontal reaction force (Fig.~\ref{fig:trapezoid_optimize_with_threshold}\deleted{(b)}\added{(c)}). This enhanced smoothness is highly advantageous for practical multi-material additive manufacturing, as it diffuses severe interfacial stress concentrations and mitigates the risk of delamination between distinct material phases during cyclic grasping operations\added{~\cite{bartlett20153d,saldivar2023bioinspired}}. \added{Although two-material topology optimization with a filter would also produce a blended interface, the response within it would rely on an assumed interpolation between the two constitutive laws, whereas here the response of every intermediate composition is given by the calibrated constitutive law, so that the optimizer accounts for the calibrated response of the graded transition.}

\subsection{Contact points: Composition optimization for anisotropic response under contact and finite deformations}
\label{sec:contact-points-large-deform}

\begin{figure}[ht]
    \centering
    \includegraphics[width=1.0\linewidth]{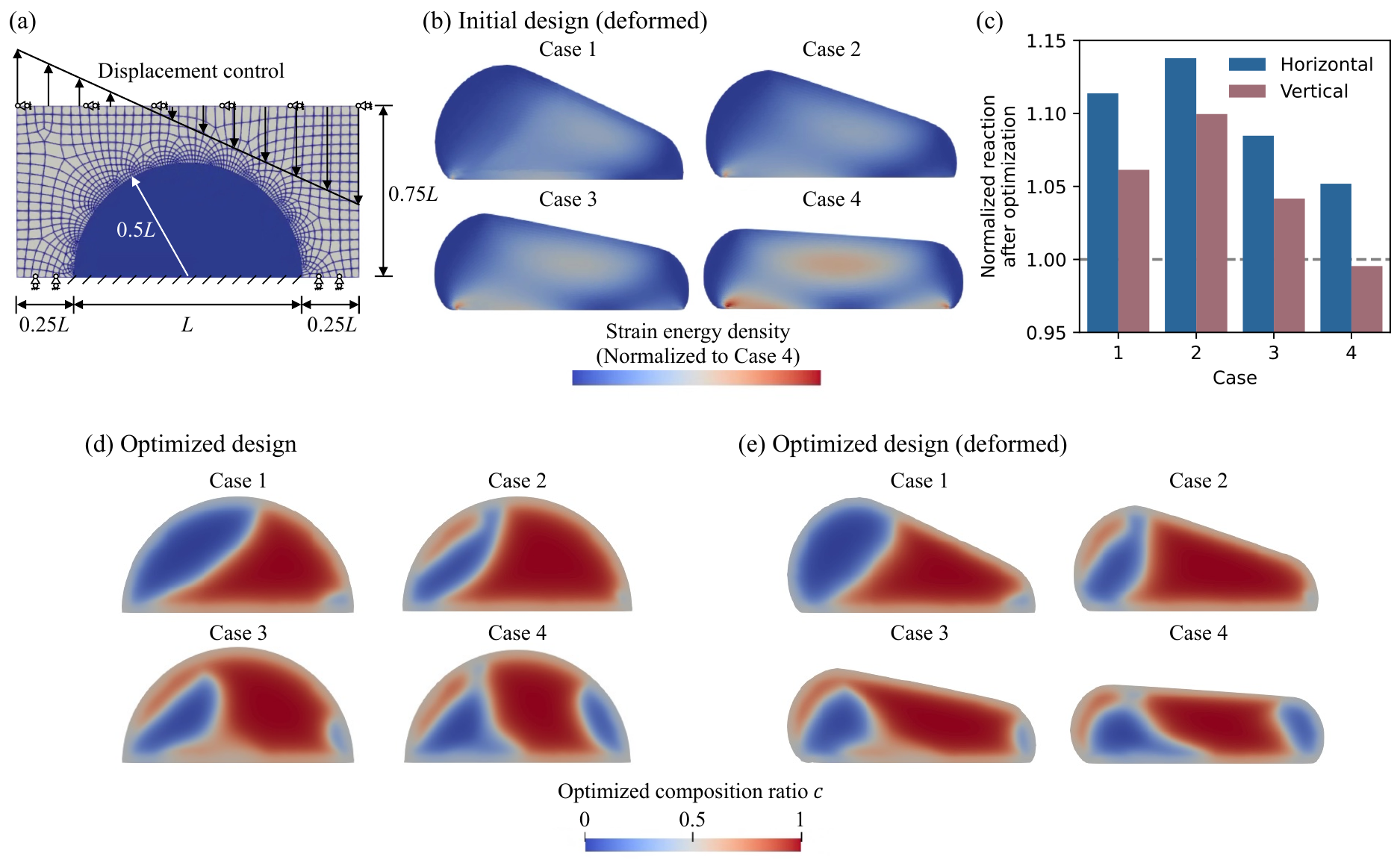}
    \caption{Optimization of the continuous composition ratio $c$ within a semicircular contact point subjected to inclined finite compression. (a) Geometry and boundary conditions of the semicircular domain and the encapsulating third medium. (b) Deformed configurations of the initial homogeneous design under various prescribed displacement profiles, illustrating the resulting strain energy density fields. (c) Horizontal and vertical reaction forces on the fixed bottom boundary after optimization, normalized with respect to the initial homogeneous design. (d) Optimized distributions of the composition ratio $c$ shown in the undeformed state. (e) Optimized designs illustrated in the deformed state. }
    \label{fig:semisphere}
\end{figure}

Extending the first contact point example to a more realistic finite deformation setup under more general contact conditions, a two-dimensional semicircle is considered, with a radius of $0.5L$ fixed at its base, subjected to compression by an inclined plate at various angles. This interaction is numerically implemented via the third medium contact approach \cite{faltus2024ThirdMediumFinite,wriggers2025ThirdMediumApproach}. As detailed in \ref{sec:third-medium-contact}, this approach utilizes a fictitious, highly compliant medium that undergoes severe deformation, and implicitly transfers the contact load to the physical body as its local volume approaches zero. In our setup, the contact point is encapsulated by this third medium. The top boundary of the third medium is subjected to a spatially varying downward displacement profile to model the inclined plate, as illustrated in Fig.~\ref{fig:semisphere}(a). The entire computational domain, including the third medium region, is discretized into 6,627 quadratic quadrilateral plane-strain elements. Note that the third medium approach necessitates at least second-order shape functions to properly compute the required higher-order spatial derivatives of the displacement field (see \ref{sec:third-medium-contact}). Similar to the previous example, the composition ratio distribution is optimized spatially in a fixed domain conforming to the shape of the contact point.

The computational procedure is divided into two phases: first, incrementally applying the prescribed displacement to the upper boundary of the third medium across 40 load steps until full contact is established; second, optimizing the composition ratio $c$ while maintaining these final boundary conditions. During the first phase, four distinct loading profiles are considered for the upper boundary $\partial \Omega_{\text{top}}$, where $x$ denotes the horizontal coordinate relative to the bottom-left corner of the domain:
\begin{equation}
    \Delta u_{\mathrm{top}}=
    \begin{cases}
    -\frac{0.7}{1.5}x, & \text{(Case 1)}\\
    -\frac{0.5}{1.5}x-0.12L, & \text{(Case 2)}\\
    -\frac{0.3}{1.5}x-0.24L, & \text{(Case 3)}\\
    -\frac{0.1}{1.5}x-0.36L, & \text{(Case 4)}
    \end{cases}
\end{equation}
The third medium contact parameters are set to $k_{3\mathrm{rd}}=10^{-5}$ and $k_{\mathrm{reg}}=10^{-3}$ \deleted{(in the unit of strain energy density)}\added{($k_{3\mathrm{rd}}$ is dimensionless and $k_{\mathrm{reg}}$ is in the unit of strain energy density times $L^2$), with the plane-strain regularization of ref.~\cite{wriggers2025ThirdMediumApproach} (\ref{sec:third-medium-contact})}. \added{The composition ratio in the third medium is kept at its initial value, $c_{3\mathrm{rd}}=0.5$.} The resulting deformed configurations of this initial homogeneous design with $c=0.5$ under the four loading cases are illustrated in Fig.~\ref{fig:semisphere}(b). 

During the second phase, the optimization problem is formulated similarly to Eq.~\eqref{eq:trapezoid_objective}, but the objective reactions are now evaluated at the fixed bottom boundary:
\begin{equation}
\begin{aligned}
\min_c\ \mathcal{J}(\mathbf{u}\added{,\rho\equiv1},c) =& 
\mathfrak{N} \underset{\mathrm{Vertical\ reaction}}{\underbrace{\left[\left |\int_{\partial\Omega_{\text{bottom}}} \mathbf{P}\cdot \mathbf{n}\cdot \mathbf{e}_y \mathrm{d}A\right|\right]}} 
- \mathfrak{N} \underset{\mathrm{Horizontal\ reaction}}{\underbrace{ \left[\left|\int_{\partial\Omega_{\text{bottom}}} \mathbf{P}\cdot \mathbf{n}\cdot \mathbf{e}_x\mathrm{d}A\right|\right]}}+1\\
\text{\deleted{$\displaystyle \mathrm{s.t.}\ f(\mathbf{u};c)=$}}\delamp\text{\deleted{$\displaystyle D_{\mathbf{u}} \biggl(\int_\Omega\psi(\mathbf{u},c)\mathrm{d}V+k_{3\mathrm{rd}}\int_{\Omega_{3\mathrm{rd}}}\psi(\mathbf{u})\mathrm{d}V$}}\delcr
\delamp\text{\deleted{$\displaystyle \qquad+k_{\mathrm{reg}}\int_{\Omega_{3\mathrm{rd}}}\psi_{\mathrm{reg}}\mathrm{d}V-\int_{\partial \Omega} \mathbf{t}\cdot \mathbf{u}\mathrm{d}A\biggr)=0.$}}\delcr
\added{\mathrm{s.t.}\ f(\mathbf{u},\rho\equiv1,c,\delta\mathbf{u})={}}&\added{D_{\mathbf{u}}\Pi(\mathbf{u},\rho\equiv1,c;\delta\mathbf{u})=0\quad\forall\,\delta\mathbf{u}\in\mathcal{V}_h,}\\
\added{\text{with}\ \Pi(\mathbf{u},\rho\equiv1,c)={}}&\added{\int_\Omega\psi(\mathbf{u},c)\mathrm{d}V+k_{3\mathrm{rd}}\int_{\Omega_{3\mathrm{rd}}}\psi(\mathbf{u},c_{3\mathrm{rd}})\mathrm{d}V}\\
&\added{\qquad+k_{\mathrm{reg}}\int_{\Omega_{3\mathrm{rd}}}\psi_{\mathrm{reg}}\mathrm{d}V-\int_{\partial(\Omega\cup\Omega_{3\mathrm{rd}})} \mathbf{t}\cdot \mathbf{u}\mathrm{d}A.}
\label{eq:semisphere_objective}
\end{aligned}
\end{equation}
The addition of 1 to the objective function is also to ensure positivity of its value. To prevent localized singularities, the underlying design variable $c$ within a $0.04L$-thick layer adjacent to the boundaries is constrained. Because the structural sensitivities are highly localized near the prescribed boundary conditions, a significantly larger initial step size of $\eta=20,000$ is employed for the Barzilai-Borwein update to accelerate convergence. The relative tolerance \texttt{TOL} (in Algorithm \ref{algo:simp}) is set to 0.01\%. A Helmholtz filter with a radius of $0.03L$ is applied to the composition ratio field $c$, and structural topology optimization remains deactivated for this example. 

Following optimization, and consistent with the trapezoidal case evaluated in the infinitesimal strain regime, the relative increase in the horizontal reaction force is significantly larger compared to the variation in the vertical reaction (Fig.~\ref{fig:semisphere}(c)). This indicates that the optimized semicircular contact point exhibits high tangential stiffness to maximize shear resistance, while maintaining high normal compliance to ensure conformal geometric adaptation. Unlike the trapezoidal strut mechanism, which could conceivably be derived via engineering intuition, the material distributions synthesized for finite inclined compression are highly non-intuitive (Fig.~\ref{fig:semisphere}(d)-(e)). This demonstrates the efficacy of the proposed end-to-end computational pipeline, revealing that algorithmic material optimization can uncover high-performance compliant architectures that elude traditional, empirical design methodologies.

\subsection{Gripper fingers: Joint optimization of material composition and topology under non-failure constraint}
\label{sec:clamping-fingers}

Topology optimization, via controlling the density field $\rho$, utilizes a high-dimensional design space that can deliver significant changes in macroscopic structural responses. Simultaneously, it can also yield unmanufacturable designs if fine features fall below the resolution limits of the fabrication process. Conversely, material composition optimization (via the composition ratio $c$) locally tunes the mechanical properties without altering the structural topology, yet it possesses an inherently more restricted design space. Therefore, concurrently optimizing both $\rho$ and $c$ offers a synergistic approach, yielding superior potential benefits compared to either optimization strategy implemented in isolation.  

Here, the design of fingers for a soft robotic gripper clamping a target object under finite deformation is investigated. Our objective is to maximize the global stiffness of the 3D-printed fingers, subject to a strict material volume constraint and a reliability constraint on the maximum permissible stretch. As demonstrated in our previous work detailing the pICNN constitutive model \cite{yangPhysicsAugmentedMachine2025} trained by experimental data of digital materials, the ultimate uniaxial stretch for regions with a low composition ratio $c$ can be up to twice that of regions with a high composition ratio. Consequently, strategically deploying softer materials facilitates adherence to the reliability constraint, albeit at the cost of reduced local stiffness. This intrinsic material tradeoff renders this example a highly practical and demonstrative scenario, requiring the computational pipeline to rigorously balance the competing demands of the objective function and the kinematic constraints. 

\begin{figure}[hp]
    \centering
    \includegraphics[width=1.0\linewidth]{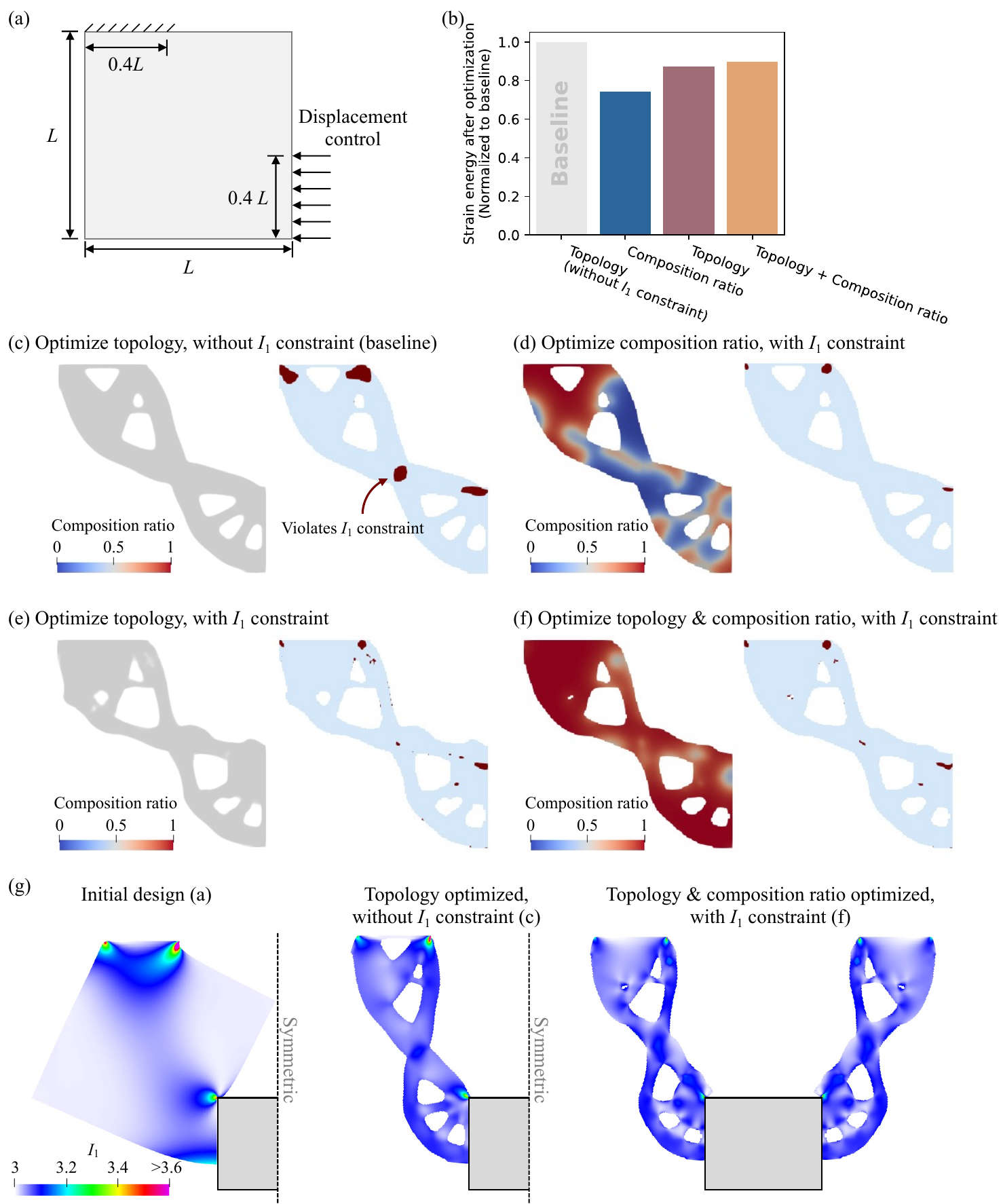}
    \caption{Optimization of the continuous composition ratio and structural topology for a soft robotic finger subjected to finite clamping deformation. (a) Geometry and boundary conditions of the initial design domain. (b) Comparison of the total strain energy (to be maximized) across the four evaluated optimization scenarios. (c)-(f) Optimization results under various strategies. Left panels: The optimized topology (void regions where $\hat{\rho}<0.5$ are masked for visualization) superimposed with the continuous composition ratio field $c$. Right panels: Spatial evaluation of the local $I_1$ failure constraint (Eq.~\eqref{eq:original-I1-constraint}), highlighting localized regions of constraint violation. (g) Deformed structural configurations of the initial homogeneous design (a), the unconstrained baseline topology (c), and the concurrently optimized topology and composition field (f), illustrated in a symmetric dual-finger grasping scenario. }
    \label{fig:clamp}
\end{figure}

As illustrated in Fig.~\ref{fig:clamp}(a), the initial design domain for the topology optimization is an $L$ by $L$ square. The leftmost $0.4L$ segment of the top boundary is fully clamped, while the bottommost $0.4L$ segment of the right boundary is subjected to a leftward prescribed displacement of $0.4L$. This displacement is applied incrementally across 10 pseudo-timesteps, after which the optimization is executed with the final boundary conditions held constant. The design domain is discretized into a structured, uniform mesh comprising 40,000 linear quadrilateral elements. A Helmholtz smoothing filter with a radius of $0.03L$ is applied to both the density and the composition ratio fields. 

The design objective is the classic stiffness maximization standard in topology optimization. For this displacement-controlled scenario, this is equivalent to maximizing the total integral of the strain energy density over the design domain. For the reliability constraint, an $I_1$-based criterion is employed, as the failure of hyperelastic polymers is predominantly governed by molecular chain stretch. The eight-chain model \cite{arruda1993three} provides a direct relationship between $I_1$ and chain stretch, which has been previously utilized to develop damage criteria \cite{mousavi2025chain}. The critical threshold $I_1^{\mathrm{crit}}$ is calibrated and computed based on the ultimate uniaxial tensile stretch limits detailed in our previous work on the pICNN model \cite{yangPhysicsAugmentedMachine2025}. For composition ratios of $c=\{0, 0.1755, 0.4669, 1.0, 2.0895\}$, the critical stretch limits are $\lambda_{\mathrm{crit}}=\{2.05, 1.80, 1.79, 1.73, 1.45\}$, which correspond to critical invariant values of $I_1^{\mathrm{crit}}=\{5.178, 4.355, 4.321, 4.149, 3.482\}$, as plotted in Fig.~\ref{fig:I_1_crit}. Through a least-squares regression, the continuous relation is obtained:
\begin{equation}
    I_1^{\mathrm{crit}}(c)=0.224c^2-1.125c+4.884.
\end{equation}

The fundamental local non-failure condition is expressed as:
\begin{equation}
\left(I_1-3\right)n_{\mathrm{safe}}\leq\left(I_{1}^{\mathrm{crit}}-3\right).
\label{eq:original-I1-constraint}
\end{equation}
This formulation is analogous to local stress constraints extensively investigated within the topology optimization community \cite{holmberg2013StressConstrainedTopology,da2021local}. Adapting this concept, the aggregated global constraint is formulated as:
\begin{equation}
    \mathfrak{N}\left[\int_{\Omega} \mathrm{ReLU}\left[\hat{\rho}^{1/\varrho}\left(I_1-3\right)n_{\mathrm{safe}}-\left(I_{1}^{\mathrm{crit}}-3\right)\right]  \mathrm{d}V\right]\leq0.25,
\end{equation}
where $n_{\mathrm{safe}}=10$ is a conservative safety factor accounting for cyclic loading conditions in robotic applications and hyperelastic degradation phenomena such as the Mullins effect. \deleted{While Ref.~\cite{holmberg2013StressConstrainedTopology} enforces pointwise stress constraints (resulting in a highly dimensional constraint space that necessitates clustering), in this work, the violation is evaluated globally into a single functional constraint.} \added{While local stress constraint approaches based on augmented Lagrangian or exterior penalty methods~\cite{da2021local,senhora2020topology} add penalty terms on the local stress violations to the objective function, the domain integral of local $I_1$ violations is here imposed as a single inequality constraint.} Furthermore, rather than strictly eliminating all local violations (which is often physically unattainable due to severe stress concentrations near the boundary conditions), a relaxed tolerance of 0.25 is permitted as it is utilized in conjunction with a strict safety factor. This effectively targets a 75\% \deleted{area} reduction \deleted{in the high-risk region} \added{in the integrated violation relative to the initial design}.  

The final optimization problem is formulated for the concurrent optimization of structural topology and material composition. To isolate and investigate the specific benefits of a spatially varying composition field, either design variable can be held constant during the optimization process. 
\begin{equation}
\begin{aligned}
\min_{\rho, c}&\ \mathcal{J}(\mathbf{u},\rho,c) = 
-\mathfrak{N}\underset{\mathrm{Stored\ energy}}{\underbrace{\left[\int_\Omega \max{\left\{\hat{\rho}^\varrho,0.001\right\}} \psi(\mathbf{u},c)\mathrm{d}V\right]}}+2,\\
\mathrm{s.t.}&\delsp\text{\deleted{$\displaystyle f(\mathbf{u};\rho,c)=D_\mathbf{u} \left(\int_\Omega \max{\left\{\hat{\rho}^\varrho,0.001\right\}}\psi_{\mathrm{interp}}(\mathbf{u},c,\mathcal{I})\mathrm{d}V-\int_{\partial \Omega} \mathbf{t}\cdot \mathbf{u}\mathrm{d}A\right)=0,$}}\delcr
\delamp\ \added{f(\mathbf{u},\rho,c,\delta\mathbf{u})=D_\mathbf{u}\Pi(\mathbf{u},\rho,c;\delta\mathbf{u})=0\quad\forall\,\delta\mathbf{u}\in\mathcal{V}_h,}\\
&\ \added{\text{with}\ \Pi(\mathbf{u},\rho,c)=\int_\Omega \max{\left\{\hat{\rho}^\varrho,0.001\right\}}\psi_{\mathrm{interp}}(\mathbf{u},c,\mathcal{I})\mathrm{d}V-\int_{\partial \Omega} \mathbf{t}\cdot \mathbf{u}\mathrm{d}A,}\\
&\ g_1(\mathbf{u},\rho,c)=\int_{\Omega} \left(\hat{\rho} - 0.35\right)  \mathrm{d}V\leq0,\\
&\ g_2(\mathbf{u},\rho,c)=\mathfrak{N}\left[\int_{\Omega} \mathrm{ReLU}\left[\hat{\rho}^{1/\varrho}\left(I_1-3\right)n_{\mathrm{safe}}-\left(I_{1}^{\mathrm{crit}}(c)-3\right)\right]  \mathrm{d}V\right]\leq0.25.
\label{eq:clamp_objective}
\end{aligned}
\end{equation}
To solve this inequality-constrained problem, the Method of Moving Asymptotes (MMA) is employed with a strict move limit of $0.01$ per iteration for both the density and the composition ratio fields. The relative tolerance \texttt{TOL} (in Algorithm \ref{algo:simp}) is set to 0.002\%. 

To facilitate a rigorous comparison demonstrating the advantages of simultaneously optimizing both structural topology and material distribution, four distinct scenarios are evaluated. First, an unconstrained topology optimization (excluding the $I_1$ reliability constraint) is performed to establish a performance baseline. This baseline design also serves as the warm-start initialization for the three subsequent constrained scenarios. In the three following scenarios, the $I_1$ constraint is introduced, and optimizations are performed for: (2) the composition ratio field only, (3) the structural topology only, and (4) the concurrent optimization of topology and composition ratio.

Fig.~\ref{fig:clamp}(b) compares the total strain energy (the stiffness objective to be maximized) across each evaluated scenario. By enforcing the $I_1$ reliability constraint, all three constrained scenarios successfully reduce the spatial area of violation compared to the unconstrained baseline (Fig.~\ref{fig:clamp}(c), which serves as the warm-start initialization), albeit by compromising the maximum attainable strain energy (right panels of Fig.~\ref{fig:clamp}(d)-(f)). While topology optimization alone (Fig.~\ref{fig:clamp}(e)) yields a higher strain energy than composition ratio optimization alone (Fig.~\ref{fig:clamp}(d)), concurrently optimizing both fields (Fig.~\ref{fig:clamp}(f)) achieves the superior global structural performance (Fig.~\ref{fig:clamp}(b)). While topology optimization drives the macroscopic structural stiffness, the composition ratio optimization locally fine-tunes the material compliance to alleviate stretch concentrations, thereby satisfying the failure constraint more efficiently. As illustrated in Fig.~\ref{fig:clamp}(g), the $I_1$ field is distributed much more uniformly throughout the concurrently optimized structure compared to the initial homogeneous design. These results indicate that the concurrent design approach mitigates localized vulnerabilities while maintaining stiffness and volume objectives.

\section{Conclusions and discussion}
\label{sec:conclusions}

\deleted{This work introduced an end-to-end computational framework for the design of multimaterial 3D-printed structures by integrating sparsified physics-augmented neural constitutive models with finite-element-based topology and material distribution optimization.} \added{This work showed that physics-augmented neural constitutive models, once sparsified into closed-form laws and combined with symbolic differentiation, enable topology and material distribution optimization of multimaterial 3D-printed structures without runtime automatic differentiation through a network or hand-derived constitutive derivatives.} By leveraging a sparsified constitutive law for digital materials, trained on physical test data, the proposed approach enables direct symbolic differentiation of stresses, tangent operators, objective functions, and adjoint sensitivities within \texttt{FEniCSx}. \deleted{This removes a key bottleneck associated with deploying standard neural constitutive models in nonlinear finite element simulations.}The resulting workflow combines data-efficient material modeling, robust nonlinear simulation, PDE-based filtering, adjoint sensitivity analysis, and constrained optimization into a unified pipeline for functionally graded digital materials with targeted structural functionality.

The framework was demonstrated on representative soft robotic gripper design problems involving tunable contact response, finite-deformation contact, and damage-aware optimization of gripper fingers. These examples show that continuous composition fields and structural topology can be optimized either independently or concurrently to achieve targeted mechanical responses while respecting local material constraints\added{, with third-medium contact and a composition-dependent failure constraint incorporated without deriving their sensitivities by hand}. More broadly, the results highlight the promise of sparse, interpretable machine-learning constitutive models as practical design primitives for multimaterial additive manufacturing. 

Future work will focus on extending the framework beyond two-dimensional quasi-static hyperelastic examples, accounting for uncertainty and manufacturability constraints that more directly reflect the resolution, mixing, and process limitations of multimaterial 3D printing. \added{In particular, the optimized designs have not been fabricated or tested in this study. The main practical obstacle is reproducing the composition field. Although the Stratasys Objet260 Connex3 printer we used to calibrate the constitutive model allows materials to be assigned voxel by voxel \cite{levin2023swelling}, it offers only the base resins and a fixed set of predefined digital materials rather than arbitrary mixing ratios. In addition, different resins deposited side by side within a layer blend over about 150~$\mu$m before curing~\cite{zorzetto2020properties}. The printing resolution, in contrast, does not restrict fabricating the present designs. The minimum member size and composition transition width are set by the filter radius $r_{\mathrm{f}}=0.03L$, equivalent to about $0.1L$ for a classical density filter \cite{lazarov2011FiltersTopologyOptimization}. For the applications we investigated, with a characteristic size $L\gtrsim15$~mm, this length scale is at least about 1.5~mm, which exceeds the voxel size $42\times42\times30~\mu$m of the Objet260 Connex3 printer and the in-layer blending length, and meets the recommended minimum wall thickness of 0.6~mm for PolyJet and of about 1.5~mm for flexible resins. Smaller components in future applications would require a larger filter radius or a minimum feature size control.}

\section*{Statements and Declarations}

\noindent \textbf{Acknowledgments.} The authors declare no funding for this work. 

\noindent \textbf{Author contributions.} Xue-Ling Luo: Methodology, Software, Investigation, Writing - Original Draft. Steven Yang: Methodology, Software. Jingye Tan: Methodology, Software. Robert F. Shepherd: Conceptualization, Writing - Review \& Editing. Noy Cohen: Conceptualization, Writing - Review \& Editing. Nikolaos Bouklas: Conceptualization, Writing - Review \& Editing, Supervision, Project administration.

\noindent \textbf{Code availability.} The developed code is available on GitHub (\url{https://github.com/LuoXueling/optimization_of_digital_material_distribution}).

\noindent \textbf{Competing Interests.} The authors declare no competing interests.

\appendix

\section{Theory of the third medium contact approach}
\label{sec:third-medium-contact}

In the numerical examples, the structures are subjected to severe mechanical contact that induces finite deformations. To efficiently model these interactions without introducing the algorithmic complexities of explicit contact mechanics, the third medium contact approach is adopted.  
The fundamental concept of the third medium approach is to introduce a fictitious, highly compliant hyperelastic domain, $\Omega_{3\mathrm{rd}}$, bridging the physical bodies $\Omega$. This medium physically prevents direct surface-to-surface interpenetration and implicitly transmits contact forces through its internal deformation. The total potential energy of the contacting bodies and the third medium is expressed as
\begin{equation}
\Pi=\int_\Omega\psi(\mathbf{u}\added{,c})\mathrm{d}V+\underset{\mathrm{Contact\ enforcement}}{\underbrace{k_{3\mathrm{rd}}\int_{\Omega_{3\mathrm{rd}}}\psi(\mathbf{u}\added{,c_{3\mathrm{rd}}})\mathrm{d}V}}
+\underset{\mathrm{Regularization}}{\underbrace{k_{\mathrm{reg}}\int_{\Omega_{3\mathrm{rd}}}\psi_{\mathrm{reg}}\mathrm{d}V}}
    -\text{\deleted{$\displaystyle\int_{\partial \Omega_{3\mathrm{rd}}} \mathbf{t}\cdot \mathbf{u}\mathrm{d}A$}}\added{\int_{\partial(\Omega\cup\Omega_{3\mathrm{rd}})} \mathbf{t}\cdot \mathbf{u}\mathrm{d}A}.
    \label{eq:third_medium_potential}
\end{equation}
\added{Here, $c_{3\mathrm{rd}}$ denotes the fixed composition ratio assigned to the third medium, which is not a design variable.} The second term in Eq.~\eqref{eq:third_medium_potential} represents the degraded strain energy of the third medium \cite{faltus2024ThirdMediumFinite}, which stiffens asymptotically as the medium's local volume approaches zero ($J \to 0$), thereby providing the necessary repulsive force to simulate contact. Beyond simulating external loading, this formulation offers significant utility in topology optimization: void regions can be naturally interpreted as the third medium, enabling the simulation of internal void closure (self-contact) without additional algorithmic overhead, although this specific feature is not activated in the present examples. When deployed within a topology optimization framework, literature suggests that the volumetric penalty term of $\psi$ (e.g., $\frac{\kappa}{2}(J-1)^2$) may be omitted for the third medium in pure plane strain contact problems \cite{faltus2024ThirdMediumFinite,wriggers2025ThirdMediumApproach}. However, to maintain consistency between the solid phase and the void (third medium) phase during material penalization, retaining this volumetric term is highly recommended \cite{faltus2024ThirdMediumFinite,bluhm2021InternalContactModeling}. 

The third term, specifically the regularization term, mitigates numerical instabilities within the third medium when $k_{3\mathrm{rd}}$ is low by penalizing severe spatial gradients in rotation and volumetric deformation. A straightforward choice of the regularization term is 
\begin{equation}
    \psi_{\mathrm{reg}}=\frac{1}{2}\mathbb{H}(\mathbf{u})\cdddot{} \mathbb{H}(\mathbf{u})=\frac{1}{2}\left \| \nabla \mathbf{F} \right \|^2.
\end{equation}
Because artificial energy stored in the third medium exacerbates non-convexity issues (such as oscillations and convergence failures), it must be minimized. Consequently, ref.~\cite{faltus2024ThirdMediumFinite} proposes incorporating only the rotational and volumetric deformations into the regularization term:
\begin{equation}
    \psi_{\mathrm{reg}}=\frac{1}{2}\left(\left \| \nabla \mathbf{R} \right \|^2 +\left\| \nabla J \right\|^2\right),
\end{equation}
and simplifies the computationally complicated gradient of the proper orthogonal rotation tensor \deleted{$\mathbf{R}=\mathbf{F}\mathbf{U}^{-1}$ (where $\mathbf{U}$ is the right stretch tensor)} \added{$\mathbf{R}=\mathbf{F}\mathbf{C}^{-1/2}$} by substituting an alternative rotational measure. This tailored regularization demonstrates superior convergence performance and more consistent mechanical behavior. Alternatively, for plane strain conditions, ref.~\cite{wriggers2025ThirdMediumApproach} proposes an even simpler formulation to account for rotational gradients:
\begin{equation}
    \psi_{\mathrm{reg}}=\frac{1}{2}\left[\left \| \nabla \left(\frac{F_{12}-F_{21}}{F_{11}+F_{22}}\right) \right \|^2 +\left\| \nabla J \right\|^2\right],
\end{equation}
where this ratio of deformation gradient components explicitly represents the tangent of the local rigid body rotation angle. Evaluating these regularization forms typically requires second-order or higher-order shape functions to adequately capture the higher-order derivatives of the displacement field. While ref.~\cite{wriggers2025ThirdMediumApproach} introduces auxiliary fields to approximate these gradients via a mixed formulation, thereby alleviating the need for high-order elements and mitigating ill-conditioning, specific meshing strategies are sufficient to ensure convergence. Specifically, (1) quadrilateral elements rather than triangular meshes are utilized to prevent excessive numerical locking and artificial stretching, and (2) larger element sizes are assigned within the third medium to restrict excessive local kinematic flexibility.

\newpage
\clearpage
\section{Supplementary figures}

\begin{figure}[h]
    \centering
    \includegraphics[width=0.5\linewidth]{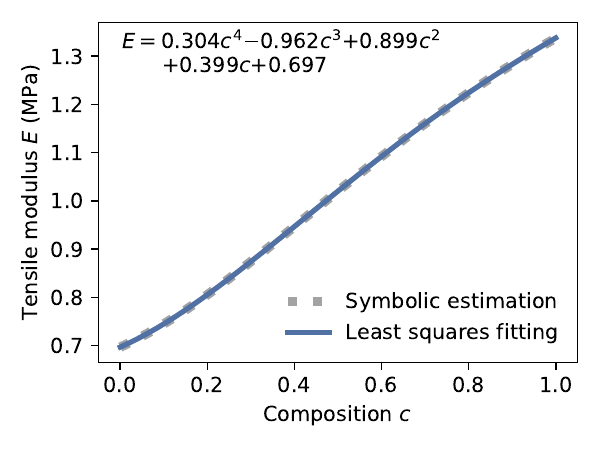}
    \caption{Symbolic derivation and estimation of tensile modulus under small strain assumption and least squares polynomial fitting.}
    \label{fig:fit_tensile_modulus}
\end{figure}

\begin{figure}[h]
    \centering
    \includegraphics[width=0.5\linewidth]{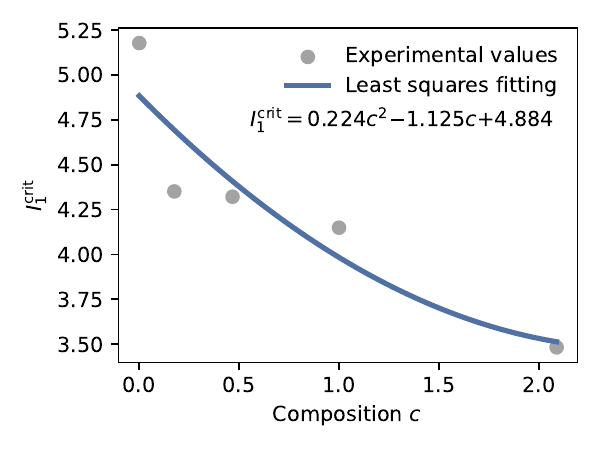}
    \caption{Critical $I_1$ values at different composition ratios and their least square fitting. }
    \label{fig:I_1_crit}
\end{figure}

\newpage
\clearpage
\section{\texorpdfstring{\added{Sensitivity analysis}}{Sensitivity analysis}}
\label{sec:appendix-sensitivity}

\added{The sensitivities of the objectives and inequality constraints in Eqs.~\eqref{eq:trapezoid_objective}, \eqref{eq:semisphere_objective}, and \eqref{eq:clamp_objective} follow from Eq.~\eqref{eq:lagrange_solution} with $\mathcal{J}$ replaced by each functional $Q\in\{\mathcal{J},g_1,g_2\}$. Since $\rho$ and $c$ enter the forms and functionals only through the filtered fields $\tilde{\rho}$ and $\tilde{c}$ (Section~\ref{sec:filter}), with the projected density $\hat{\rho}=\hat{\rho}(\tilde{\rho})$ and with the void indicator $\mathcal{I}$ held fixed as stated below, the derivatives are taken with respect to $\tilde{\rho}$ and $\tilde{c}$, and the filter chain rule at the end maps them to the underlying fields. For each $Q$, Eqs.~\eqref{eq:lagrange_solution}a and b give the adjoint $\boldsymbol{\lambda}_Q\in\mathcal{V}_h$ and the sensitivity, i.e., the derivative of $\mathcal{L}_Q(\mathbf{u},\rho,c,\boldsymbol{\lambda}_Q)=Q(\mathbf{u},\rho,c)+f(\mathbf{u},\rho,c,\boldsymbol{\lambda}_Q)$ with respect to $\tilde{v}\in\{\tilde{\rho},\tilde{c}\}$,}
\begin{equation}
\begin{aligned}
    \added{D_\mathbf{u}f(\mathbf{u},\rho,c,\boldsymbol{\lambda}_Q;\delta\mathbf{u})}&\added{{}=-D_\mathbf{u}Q(\mathbf{u},\rho,c;\delta\mathbf{u})\qquad\forall\,\delta\mathbf{u}\in\mathcal{V}_h,}\\
    \added{D_{\tilde{v}}\mathcal{L}_Q(\mathbf{u},\rho,c,\boldsymbol{\lambda}_Q;\delta\tilde{v})}&\added{{}=D_{\tilde{v}}Q(\mathbf{u},\rho,c;\delta\tilde{v})+D_{\tilde{v}}f(\mathbf{u},\rho,c,\boldsymbol{\lambda}_Q;\delta\tilde{v}),}
\end{aligned}
\label{eq:sens-adjoint}
\end{equation}
\added{where $D_\mathbf{u}f(\mathbf{u},\rho,c,\boldsymbol{\lambda}_Q;\delta\mathbf{u})$ is the form of the symmetric Jacobian of the forward problem and is omitted below. In this work, each required term, except the filter chain rule below, is obtained by automatic symbolic differentiation in \texttt{FEniCSx}, and the explicit expressions are presented solely for reproducibility. }

\added{Below, the terms required by Eq.~\eqref{eq:sens-adjoint} are listed for each example, followed by the expressions of the terms left unexpanded. The penalization factor $\varrho$, the normalization constants of $\mathfrak{N}[\cdot]$, i.e., the initial values $R_{k,0}$, $\mathcal{W}_0$, and $\mathcal{G}_0$ below, and the void indicator $\mathcal{I}$ (Section~\ref{sec:simp}) are held fixed. }

\added{\textbf{Contact points} (Eqs.~\eqref{eq:trapezoid_objective} and \eqref{eq:semisphere_objective}). Only $c$ is optimized, and the residual is}
\begin{equation}
    \added{f(\mathbf{u},\rho\equiv1,c,\delta\mathbf{u})=\int_\Omega\mathbf{P}(\mathbf{F},\tilde{c}):\nabla\delta\mathbf{u}\,\mathrm{d}V+f_0(\mathbf{u},\delta\mathbf{u}),}
\end{equation}
\added{where $f_0$ collects the design-independent terms, i.e., the traction term and, in Eq.~\eqref{eq:semisphere_objective}, the third medium terms (with fixed $c_{3\mathrm{rd}}$), so that}
\begin{equation}
    \added{D_{\tilde{c}}f(\mathbf{u},\rho\equiv1,c,\boldsymbol{\lambda}_Q;\delta\tilde{c})=\int_\Omega\delta\tilde{c}\ \frac{\partial\mathbf{P}}{\partial c}:\nabla\boldsymbol{\lambda}_Q\,\mathrm{d}V.}
    \label{eq:sens-contact-f}
\end{equation}
\added{With the reactions $R_k=\int_\Gamma(\mathbf{P}\mathbf{n})\cdot\mathbf{e}_k\,\mathrm{d}A$ on $\Gamma=\partial\Omega_{\mathrm{top}}$ (Eq.~\eqref{eq:trapezoid_objective}) or $\partial\Omega_{\mathrm{bottom}}$ (Eq.~\eqref{eq:semisphere_objective}), and $\mathbf{a}=\mathrm{sgn}(R_y)\mathbf{e}_y/|R_{y,0}|-\mathrm{sgn}(R_x)\mathbf{e}_x/|R_{x,0}|$, the objective $\mathcal{J}$ gives}
\begin{equation}
\begin{aligned}
    \added{D_\mathbf{u}\mathcal{J}(\mathbf{u},\rho\equiv1,c;\delta\mathbf{u})}&\added{{}=\int_\Gamma\left(\left(\frac{\partial\mathbf{P}}{\partial\mathbf{F}}:\nabla\delta\mathbf{u}\right)\mathbf{n}\right)\cdot\mathbf{a}\,\mathrm{d}A,}\\
    \added{D_{\tilde{c}}\mathcal{J}(\mathbf{u},\rho\equiv1,c;\delta\tilde{c})}&\added{{}=\int_\Gamma\delta\tilde{c}\left(\frac{\partial\mathbf{P}}{\partial c}\mathbf{n}\right)\cdot\mathbf{a}\,\mathrm{d}A.}
\end{aligned}
\label{eq:sens-contact}
\end{equation}

\added{\textbf{Gripper fingers} (Eq.~\eqref{eq:clamp_objective}). With the stiffness interpolation $\phi=\max\{\hat{\rho}^{\varrho},0.001\}$, the residual is}
\begin{equation}
    \added{f(\mathbf{u},\rho,c,\delta\mathbf{u})=\int_\Omega\phi(\hat{\rho})\,\mathbf{P}_{\mathrm{interp}}(\mathbf{F},\tilde{c}):\nabla\delta\mathbf{u}\,\mathrm{d}V-\int_{\partial\Omega}\mathbf{t}\cdot\delta\mathbf{u}\,\mathrm{d}A,}
\end{equation}
\added{so that}
\begin{equation}
\begin{aligned}
    \added{D_{\tilde{\rho}}f(\mathbf{u},\rho,c,\boldsymbol{\lambda}_Q;\delta\tilde{\rho})}&\added{{}=\int_\Omega\phi'\hat{\rho}'\,\delta\tilde{\rho}\ \mathbf{P}_{\mathrm{interp}}:\nabla\boldsymbol{\lambda}_Q\,\mathrm{d}V,}\\
    \added{D_{\tilde{c}}f(\mathbf{u},\rho,c,\boldsymbol{\lambda}_Q;\delta\tilde{c})}&\added{{}=\int_\Omega\phi\,\delta\tilde{c}\ \frac{\partial\mathbf{P}_{\mathrm{interp}}}{\partial c}:\nabla\boldsymbol{\lambda}_Q\,\mathrm{d}V.}
\end{aligned}
\label{eq:sens-clamp-f}
\end{equation}
\added{When only $c$ is optimized, $\phi$ is evaluated at the fixed $\hat{\rho}$ and only the $\tilde{c}$-sensitivities are used. For the objective $\mathcal{J}=-\mathcal{W}/\mathcal{W}_0+2$ with $\mathcal{W}=\int_\Omega\phi\,\psi(\mathbf{F},\tilde{c})\,\mathrm{d}V$ (the stored energy uses $\psi$ rather than $\psi_{\mathrm{interp}}$),}
\begin{equation}
\begin{aligned}
    \added{D_\mathbf{u}\mathcal{J}(\mathbf{u},\rho,c;\delta\mathbf{u})}&\added{{}=-\frac{1}{\mathcal{W}_0}\int_\Omega\phi\,\mathbf{P}:\nabla\delta\mathbf{u}\,\mathrm{d}V,}\\
    \added{D_{\tilde{\rho}}\mathcal{J}(\mathbf{u},\rho,c;\delta\tilde{\rho})}&\added{{}=-\frac{1}{\mathcal{W}_0}\int_\Omega\phi'\hat{\rho}'\,\delta\tilde{\rho}\ \psi\,\mathrm{d}V,}\\
    \added{D_{\tilde{c}}\mathcal{J}(\mathbf{u},\rho,c;\delta\tilde{c})}&\added{{}=-\frac{1}{\mathcal{W}_0}\int_\Omega\phi\,\delta\tilde{c}\ \frac{\partial\psi}{\partial c}\,\mathrm{d}V.}
\end{aligned}
\label{eq:sens-clamp-J}
\end{equation}
\added{The volume constraint depends only on $\hat{\rho}$, so that $\boldsymbol{\lambda}_{g_1}=\mathbf{0}$ and}
\begin{equation}
\begin{aligned}
    \added{D_\mathbf{u}g_1(\mathbf{u},\rho,c;\delta\mathbf{u})}&\added{{}=0,\qquad D_{\tilde{c}}g_1(\mathbf{u},\rho,c;\delta\tilde{c})=0,}\\
    \added{D_{\tilde{\rho}}g_1(\mathbf{u},\rho,c;\delta\tilde{\rho})}&\added{{}=\int_\Omega\hat{\rho}'\,\delta\tilde{\rho}\,\mathrm{d}V.}
\end{aligned}
\label{eq:sens-clamp-g1}
\end{equation}
\added{For $g_2=\mathcal{G}/\mathcal{G}_0$ with $\mathcal{G}=\int_\Omega\mathrm{ReLU}(\xi)\,\mathrm{d}V$, $\xi=\hat{\rho}^{1/\varrho}(I_1-3)\,n_{\mathrm{safe}}-(I_1^{\mathrm{crit}}(\tilde{c})-3)$, and $\mathcal{H}=\mathrm{ReLU}'$,}
\begin{equation}
\begin{aligned}
    \added{D_\mathbf{u}g_2(\mathbf{u},\rho,c;\delta\mathbf{u})}&\added{{}=\frac{2n_{\mathrm{safe}}}{\mathcal{G}_0}\int_\Omega\mathcal{H}(\xi)\,\hat{\rho}^{1/\varrho}\,\mathbf{F}:\nabla\delta\mathbf{u}\,\mathrm{d}V,}\\
    \added{D_{\tilde{\rho}}g_2(\mathbf{u},\rho,c;\delta\tilde{\rho})}&\added{{}=\frac{n_{\mathrm{safe}}}{\varrho\,\mathcal{G}_0}\int_\Omega\hat{\rho}'\,\delta\tilde{\rho}\ \mathcal{H}(\xi)\,\hat{\rho}^{1/\varrho-1}(I_1-3)\,\mathrm{d}V,}\\
    \added{D_{\tilde{c}}g_2(\mathbf{u},\rho,c;\delta\tilde{c})}&\added{{}=-\frac{1}{\mathcal{G}_0}\int_\Omega\delta\tilde{c}\ \mathcal{H}(\xi)\,\frac{\mathrm{d}I_1^{\mathrm{crit}}}{\mathrm{d}c}\,\mathrm{d}V.}
\end{aligned}
\label{eq:sens-clamp-g2}
\end{equation}

\added{\textbf{Derivative terms.} The projection of Eq.~\eqref{eq:density-projection} and the stiffness interpolation give}
\begin{equation}
    \added{\hat{\rho}'=\frac{\beta\left[1-\tanh^2\left(\beta(\tilde{\rho}-\tau)\right)\right]}{\tanh(\beta\tau)+\tanh\left(\beta(1-\tau)\right)},\qquad}
    \added{\phi'=}\begin{cases}\added{\varrho\,\hat{\rho}^{\varrho-1},} & \added{\hat{\rho}^{\varrho}\geq0.001,}\\ \added{0,} & \added{\text{otherwise}.}\end{cases}
\end{equation}
\added{Let $\psi^{NN}_{,i}=\partial\psi^{NN}/\partial\bar{I}_i$ and $\psi^{NN}_{,ic}=\partial\psi^{NN}_{,i}/\partial c$ (Eq.~\eqref{eq:psi-NN}), $I_2=\frac{1}{2}[I_1^2-\mathrm{tr}\,\mathbf{C}^2]$, and $\mathbf{G}_i=\partial\bar{I}_i/\partial\mathbf{F}$, with the plane strain $\mathbf{F}$ embedded in 3D ($F_{33}=1$):}
\begin{equation}
    \added{\mathbf{G}_1=2J^{-2/3}\left(\mathbf{F}-\frac{I_1}{3}\mathbf{F}^{-\mathrm{T}}\right),\qquad}
    \added{\mathbf{G}_2=2J^{-4/3}\left(I_1\mathbf{F}-\mathbf{F}\mathbf{C}-\frac{2I_2}{3}\mathbf{F}^{-\mathrm{T}}\right).}
\end{equation}
\added{The model in Eq.~\eqref{eq:total_strain_energy_NN_soft} gives}
\begin{equation}
\begin{aligned}
    \added{\mathbf{P}}&\added{{}=\psi^{NN}_{,1}\mathbf{G}_1+\psi^{NN}_{,2}\mathbf{G}_2+\kappa(J-1)J\mathbf{F}^{-\mathrm{T}},}\\
    \added{\frac{\partial\mathbf{P}}{\partial c}}&\added{{}=\psi^{NN}_{,1c}\mathbf{G}_1+\psi^{NN}_{,2c}\mathbf{G}_2+\kappa'(J-1)J\mathbf{F}^{-\mathrm{T}},}\\
    \added{\frac{\partial\psi}{\partial c}}&\added{{}=\frac{\partial\psi^{NN}}{\partial c}\Big|_{(\bar{I}_1,\bar{I}_2)}-\frac{\partial\psi^{NN}}{\partial c}\Big|_{(3,3)}+\frac{1}{2}\kappa'(J-1)^2,}
\end{aligned}
\end{equation}
\added{with $\kappa'=E'/[3(1-2\nu)]$ and $E'=\mathrm{d}E/\mathrm{d}c$ from Eq.~\eqref{eq:fitted-E-c}; $\partial\mathbf{P}/\partial\mathbf{F}$ is the tangent of the forward problem. For the void interpolation, with $\mathbf{F}_{\mathcal{I}}=\mathbf{I}+\mathcal{I}(\mathbf{F}-\mathbf{I})$, $\mu=E/[2(1+\nu)]$, and $\mu'=E'/[2(1+\nu)]$,}
\begin{equation}
    \added{\mathbf{P}_{\mathrm{interp}}=\mathcal{I}\left[\mathbf{P}(\mathbf{F}_{\mathcal{I}},\tilde{c})-\mathbf{P}_L(\mathbf{F}_{\mathcal{I}},\tilde{c})\right]+\mathbf{P}_L(\mathbf{F},\tilde{c}),\qquad}
    \added{\mathbf{P}_L=\frac{\mu}{2}\mathbf{G}_1+\kappa(J-1)J\mathbf{F}^{-\mathrm{T}},}
    \label{eq:sens-Pinterp}
\end{equation}
\added{and $\partial\mathbf{P}_{\mathrm{interp}}/\partial c$ follows from Eq.~\eqref{eq:sens-Pinterp} with $\mathbf{P}\to\partial\mathbf{P}/\partial c$, $\mu\to\mu'$, and $\kappa\to\kappa'$.}

\added{\textbf{Filter chain rule.} The gradients with respect to the nodal vectors of the underlying fields follow from those with respect to the filtered fields by the filter transpose (Section~\ref{sec:filter}),}
\begin{equation}
    \added{\nabla_{\mathbf{v}}\mathcal{L}_Q=\mathbf{M}_{\mathrm{f}}\mathbf{K}_{\mathrm{f}}^{-1}\nabla_{\tilde{\mathbf{v}}}\mathcal{L}_Q,\qquad}
    \added{(\nabla_{\tilde{\mathbf{v}}}\mathcal{L}_Q)_k=D_{\tilde{v}}\mathcal{L}_Q(\mathbf{u},\rho,c,\boldsymbol{\lambda}_Q;N^v_k),\qquad v\in\{\rho,c\}.}
    \label{eq:sens-filter-chain}
\end{equation}


\clearpage
\newpage

\begin{thebibliography}{10}
\expandafter\ifx\csname url\endcsname\relax
  \def\url#1{\texttt{#1}}\fi
\expandafter\ifx\csname urlprefix\endcsname\relax\def\urlprefix{URL }\fi
\expandafter\ifx\csname href\endcsname\relax
  \def\href#1#2{#2} \def\path#1{#1}\fi

\bibitem{ahn20243DPrinting3D}
S.-J. Ahn, H.~Lee, K.-J. Cho, {{3D}} printing with a {{3D}} printed digital material filament for programming functional gradients, Nature Communications 15~(1) (2024) 3605.
\newblock \href {https://doi.org/10.1038/s41467-024-47480-5} {\path{doi:10.1038/s41467-024-47480-5}}.

\bibitem{rossetti2017microstructure}
L.~Rossetti, L.~Kuntz, E.~Kunold, J.~Schock, K.~M{\"u}ller, H.~Grabmayr, J.~Stolberg-Stolberg, F.~Pfeiffer, S.~Sieber, R.~Burgkart, et~al., The microstructure and micromechanics of the tendon--bone insertion, Nature materials 16~(6) (2017) 664--670.

\bibitem{pragya2023SoftFunctionallyGradient}
A.~Pragya, T.~K. Ghosh, Soft {{Functionally Gradient Materials}} and {{Structures}} -- {{Natural}} and {{Manmade}}: {{A Review}}, Advanced Materials 35~(49) (2023) 2300912.
\newblock \href {https://doi.org/10.1002/adma.202300912} {\path{doi:10.1002/adma.202300912}}.

\bibitem{bartlett20153d}
N.~W. Bartlett, M.~T. Tolley, J.~T. Overvelde, J.~C. Weaver, B.~Mosadegh, K.~Bertoldi, G.~M. Whitesides, R.~J. Wood, A 3d-printed, functionally graded soft robot powered by combustion, Science 349~(6244) (2015) 161--165.

\bibitem{kumar2013development}
S.~Kumar, K.~M. Reddy, A.~Kumar, G.~R. Devi, Development and characterization of polymer--ceramic continuous fiber reinforced functionally graded composites for aerospace application, Aerospace Science and Technology 26~(1) (2013) 185--191.

\bibitem{saleh202030}
B.~Saleh, J.~Jiang, R.~Fathi, T.~Al-Hababi, Q.~Xu, L.~Wang, D.~Song, A.~Ma, 30 years of functionally graded materials: An overview of manufacturing methods, applications and future challenges, Composites Part B: Engineering 201 (2020) 108376.

\bibitem{li2025high}
K.~Li, M.~Zhang, Z.~Zhang, P.~Jin, Y.~Wang, W.~Yan, L.~Zhu, D.~Z. Zhang, L.~E. Murr, High performance realization of functionally graded materials based on integrated optimal design and additive manufacturing: A review, International Materials Reviews 70~(6) (2025) 497--547.

\bibitem{muthuram2022review}
N.~Muthuram, P.~S. Madhav, D.~K. Vasan, M.~E. Mohan, G.~Prajeeth, A review of recent literatures in poly jet printing process, Materials Today: Proceedings 68 (2022) 1906--1920.

\bibitem{wei2019experimental}
X.~Wei, A.~Bhardwaj, C.-C. Shih, L.~Zeng, B.~Tai, Z.~Pei, Experimental investigation of stratasys j750 polyjet printer: Effects of orientation and layer thickness on mechanical properties, in: International Manufacturing Science and Engineering Conference, Vol. 58745, American Society of Mechanical Engineers, 2019, p. V001T01A021.

\bibitem{levin2023swelling}
M.~Levin, N.~Cohen, Swelling under constraints: exploiting 3d-printing to optimize the performance of gel-based devices, Advanced Materials Technologies 8~(14) (2023) 2202136.

\bibitem{wade2025implicit}
C.~Wade, D.~Beck, R.~MacCurdy, Implicit modeling for 3d-printed multi-material computational object design via python, in: Proceedings of the ACM Symposium on Computational Fabrication, 2025, pp. 1--16.

\bibitem{smith2021seamless}
L.~Smith, T.~Hainsworth, Z.~Jordan, X.~Bell, R.~MacCurdy, A seamless workflow for design and fabrication of multimaterial pneumatic soft actuators, in: 2021 IEEE 17th International Conference on Automation Science and Engineering (CASE), IEEE, 2021, pp. 718--723.

\bibitem{nazir2023multi}
A.~Nazir, O.~Gokcekaya, K.~M.~M. Billah, O.~Ertugrul, J.~Jiang, J.~Sun, S.~Hussain, Multi-material additive manufacturing: A systematic review of design, properties, applications, challenges, and 3d printing of materials and cellular metamaterials, Materials \& Design 226 (2023) 111661.

\bibitem{hirano2025modeling}
T.~Hirano, Modeling and design of micro-structures: focusing on functionally graded materials and future prospects, Frontiers in Materials 12 (2025) 1659727.

\bibitem{garg2026machine}
A.~Garg, W.~Zheng, R.~Raman, L.~Li, Machine learning in functionally graded materials and nano fgms: A comprehensive review of predictive modeling for mechanical behavior: A. garg et al., Archives of Computational Methods in Engineering 33~(1) (2026) 533--575.

\bibitem{bocciarelli2008constitutive}
M.~Bocciarelli, G.~Bolzon, G.~Maier, A constitutive model of metal--ceramic functionally graded material behavior: formulation and parameter identification, Computational Materials Science 43~(1) (2008) 16--26.

\bibitem{pascon2018large}
J.~P. Pascon, Large deformation analysis of functionally graded visco-hyperelastic materials, Computers \& Structures 206 (2018) 90--108.

\bibitem{xue2014phenomenological}
L.~Xue, G.~Dui, B.~Liu, L.~Xin, A phenomenological constitutive model for functionally graded porous shape memory alloy, International journal of engineering science 78 (2014) 103--113.

\bibitem{hornik1989multilayer}
K.~Hornik, M.~Stinchcombe, H.~White, Multilayer feedforward networks are universal approximators, Neural networks 2~(5) (1989) 359--366.

\bibitem{lu2019deeponet}
L.~Lu, P.~Jin, G.~E. Karniadakis, Deeponet: Learning nonlinear operators for identifying differential equations based on the universal approximation theorem of operators, arXiv preprint arXiv:1910.03193 (2019).

\bibitem{fuhg2025ReviewDataDrivenConstitutive}
J.~N. Fuhg, G.~Anantha~Padmanabha, N.~Bouklas, B.~Bahmani, W.~Sun, N.~N. Vlassis, M.~Flaschel, P.~Carrara, L.~De~Lorenzis, A {{Review}} on {{Data-Driven Constitutive Laws}} for {{Solids}}, Archives of Computational Methods in Engineering 32~(3) (2025) 1841--1883.
\newblock \href {https://doi.org/10.1007/s11831-024-10196-2} {\path{doi:10.1007/s11831-024-10196-2}}.

\bibitem{kirchdoerferDatadrivenComputationalMechanics2016}
T.~Kirchdoerfer, M.~Ortiz, Data-driven computational mechanics, Computer Methods in Applied Mechanics and Engineering 304 (2016) 81--101.
\newblock \href {https://doi.org/10.1016/j.cma.2016.02.001} {\path{doi:10.1016/j.cma.2016.02.001}}.

\bibitem{tangMAP123DatadrivenApproach2019}
S.~Tang, G.~Zhang, H.~Yang, Y.~Li, W.~K. Liu, X.~Guo, {{MAP123}}: {{A}} data-driven approach to use {{1D}} data for {{3D}} nonlinear elastic materials modeling, Computer Methods in Applied Mechanics and Engineering 357 (2019) 112587.
\newblock \href {https://doi.org/10.1016/j.cma.2019.112587} {\path{doi:10.1016/j.cma.2019.112587}}.

\bibitem{luo2022data}
X.-L. Luo, J.-Y. Ye, P.-S. Ma, L.-W. Zhang, Data-driven enhanced phase field models for highly accurate prediction of mode i and mode ii fracture, Computer Methods in Applied Mechanics and Engineering 400 (2022) 115535.

\bibitem{guoDatadrivenApproachPredicting2022}
H.-J. Guo, A data-driven approach to predicting the anisotropic mechanical behaviour of voided single crystals, Journal of the Mechanics and Physics of Solids (2022).

\bibitem{klein2022polyconvex}
D.~K. Klein, M.~Fern{\'a}ndez, R.~J. Martin, P.~Neff, O.~Weeger, Polyconvex anisotropic hyperelasticity with neural networks, Journal of the Mechanics and Physics of Solids 159 (2022) 104703.

\bibitem{xuLearningConstitutiveRelations2021}
K.~Xu, D.~Z. Huang, E.~Darve, Learning constitutive relations using symmetric positive definite neural networks, Journal of Computational Physics 428 (2021) 110072.
\newblock \href {https://doi.org/10.1016/j.jcp.2020.110072} {\path{doi:10.1016/j.jcp.2020.110072}}.

\bibitem{linka2021constitutive}
K.~Linka, M.~Hillg{\"a}rtner, K.~P. Abdolazizi, R.~C. Aydin, M.~Itskov, C.~J. Cyron, Constitutive artificial neural networks: A fast and general approach to predictive data-driven constitutive modeling by deep learning, Journal of Computational Physics 429 (2021) 110010.

\bibitem{ciftciPhysicsinformedGANFramework2024}
K.~Ciftci, K.~Hackl, A physics-informed {{GAN}} framework based on model-free data-driven computational mechanics, Computer Methods in Applied Mechanics and Engineering 424 (2024) 116907.
\newblock \href {https://doi.org/10.1016/j.cma.2024.116907} {\path{doi:10.1016/j.cma.2024.116907}}.

\bibitem{fuhg2022physics}
J.~N. Fuhg, N.~Bouklas, On physics-informed data-driven isotropic and anisotropic constitutive models through probabilistic machine learning and space-filling sampling, Computer Methods in Applied Mechanics and Engineering 394 (2022) 114915.

\bibitem{fuhg2024stress}
J.~N. Fuhg, N.~Bouklas, R.~E. Jones, Stress representations for tensor basis neural networks: alternative formulations to finger--rivlin--ericksen, Journal of Computing and Information Science in Engineering 24~(11) (2024) 111007.

\bibitem{tac2022data}
V.~Tac, F.~S. Costabal, A.~B. Tepole, Data-driven tissue mechanics with polyconvex neural ordinary differential equations, Computer Methods in Applied Mechanics and Engineering 398 (2022) 115248.

\bibitem{luo2025physics}
X.-L. Luo, C.-C. Lyu, L.-W. Zhang, Physics-informed ensemble learning for robustly extrapolating and revealing fatigue life of composites, Composites Science and Technology (2025) 111302.

\bibitem{janssen2024physics}
J.~A. Janssen, G.~Haikal, E.~C. DeCarlo, M.~J. Hartnett, M.~L. Kirby, A physics-informed general convolutional network for the computational modeling of materials with damage, Journal of Computing and Information Science in Engineering 24~(11) (2024) 111002.

\bibitem{luo2026physics}
X.~Luo, J.~Huang, L.~Shen, H.~Wang, Z.~Shi, Physics-informed neural networks for constitutive modeling and multiphysics coupling in viscoelastic materials: Applications to asphalt pavement mechanics, Neural Networks (2026) 108803.

\bibitem{garbrecht2021InterpretableMachineLearning}
K.~Garbrecht, M.~Aguilo, A.~Sanderson, A.~Rollett, R.~M. Kirby, J.~Hochhalter, Interpretable {{Machine Learning}} for {{Texture-Dependent Constitutive Models}} with {{Automatic Code Generation}} for {{Topological Optimization}}, Integrating Materials and Manufacturing Innovation 10~(3) (2021) 373--392.
\newblock \href {https://doi.org/10.1007/s40192-021-00231-6} {\path{doi:10.1007/s40192-021-00231-6}}.

\bibitem{yangPhysicsAugmentedMachine2025}
S.~Yang, M.~Levin, G.~Anantha~Padmanabha, M.~Borshevsky, O.~Cohen, D.~T. Seidl, R.~E. Jones, N.~Bouklas, N.~Cohen, Physics augmented machine learning discovery of composition-dependent constitutive laws for {{3D}} printed digital materials, International Journal of Engineering Science 217 (2025) 104381.
\newblock \href {https://doi.org/10.1016/j.ijengsci.2025.104381} {\path{doi:10.1016/j.ijengsci.2025.104381}}.

\bibitem{revels2016forward}
J.~Revels, M.~Lubin, T.~Papamarkou, Forward-mode automatic differentiation in julia, arXiv preprint arXiv:1607.07892 (2016).

\bibitem{knoll2004jacobian}
D.~A. Knoll, D.~E. Keyes, Jacobian-free newton--krylov methods: a survey of approaches and applications, Journal of Computational Physics 193~(2) (2004) 357--397.

\bibitem{alheit2026commet}
B.~Alheit, M.~Peirlinck, S.~Kumar, Commet: Orders-of-magnitude speed-up in finite element method via batch-vectorized neural constitutive updates, Computer Methods in Applied Mechanics and Engineering 452 (2026) 118728.

\bibitem{brunton2016discovering}
S.~L. Brunton, J.~L. Proctor, J.~N. Kutz, Discovering governing equations from data by sparse identification of nonlinear dynamical systems, Proceedings of the national academy of sciences 113~(15) (2016) 3932--3937.

\bibitem{Baratta_DOLFINx_the_next_2023}
I.~A. Baratta, J.~P. Dean, J.~S. Dokken, M.~Habera, J.~S. Hale, C.~N. Richardson, M.~E. Rognes, M.~W. Scroggs, N.~Sime, G.~N. Wells, {DOLFINx: the next generation FEniCS problem solving environment}, preprint (2023).
\newblock \href {https://doi.org/10.5281/zenodo.10447666} {\path{doi:10.5281/zenodo.10447666}}.

\bibitem{flaschel2021unsupervised}
M.~Flaschel, S.~Kumar, L.~De~Lorenzis, Unsupervised discovery of interpretable hyperelastic constitutive laws, Computer Methods in Applied Mechanics and Engineering 381 (2021) 113852.

\bibitem{fuhg2024extreme}
J.~N. Fuhg, R.~E. Jones, N.~Bouklas, Extreme sparsification of physics-augmented neural networks for interpretable model discovery in mechanics, Computer Methods in Applied Mechanics and Engineering 426 (2024) 116973.

\bibitem{sigmund2013topology}
O.~Sigmund, K.~Maute, Topology optimization approaches: A comparative review, Structural and multidisciplinary optimization 48~(6) (2013) 1031--1055.

\bibitem{yu2025multi}
X.~Yu, J.~Griffis, G.~Manogharan, A.~Panesar, Multi-material additive manufacturing: a computational design perspective, Virtual and Physical Prototyping 20~(1) (2025) e2546671.

\bibitem{li2025multi}
Z.~Li, H.~Xu, S.~Zhang, J.~Cui, X.~Liu, Multi-material structures topology optimization for thin-walled tube used by vehicles under static load: A review: Z. li et al., Archives of Computational Methods in Engineering 32~(8) (2025) 4769--4809.

\bibitem{zhang2018multi}
X.~S. Zhang, G.~H. Paulino, A.~S. Ramos~Jr, Multi-material topology optimization with multiple volume constraints: a general approach applied to ground structures with material nonlinearity, Structural and Multidisciplinary Optimization 57~(1) (2018) 161--182.

\bibitem{zuo2017multi}
W.~Zuo, K.~Saitou, Multi-material topology optimization using ordered simp interpolation, Structural and Multidisciplinary Optimization 55~(2) (2017) 477--491.

\bibitem{meng2025multi}
Z.~Meng, Y.~Ren, H.~Tang, Multi-material topology optimization design of microstructures with extreme mechanical properties, Applied Mathematical Modelling 140 (2025) 115895.

\bibitem{sarkar2026generalized}
S.~I. Sarkar, N.~Singh, A.~Saxena, P.~Kumar, A generalized shape function approach for multimaterial topology optimization, International Journal for Numerical Methods in Engineering 127~(16) (2026) e70401.

\bibitem{zhang2020adaptive}
X.~S. Zhang, H.~Chi, G.~H. Paulino, Adaptive multi-material topology optimization with hyperelastic materials under large deformations: A virtual element approach, Computer Methods in Applied Mechanics and Engineering 370 (2020) 112976.

\bibitem{song2026neo}
L.~Song, F.~Wang, T.~Gao, W.~Zhang, A neo-yeoh hyperelastic interpolation model for stable multi-material topology optimization under geometric nonlinearity, Computer Methods in Applied Mechanics and Engineering 448 (2026) 118467.

\bibitem{silva2024minimizing}
R.~F. Silva, P.~G. Coelho, F.~M. Conde, B.~R. Santos, J.~P. Oliveira, Minimizing the maximum von mises stress of elastic continuum structures using topology optimization and additively manufactured functionally graded materials, Computers \& Structures 301 (2024) 107469.

\bibitem{valizadeh2025minimizing}
I.~Valizadeh, J.~O. Schommartz, O.~Weeger, Minimizing stress concentrations through material optimization with grayscale vat photopolymerization, Progress in Additive Manufacturing 10~(12) (2025) 11979--11993.

\bibitem{saldivar2023bioinspired}
M.~C. Sald{\'\i}var, E.~Tay, A.~Isaakidou, V.~Moosabeiki, L.~E. Fratila-Apachitei, E.~L. Doubrovski, M.~J. Mirzaali, A.~A. Zadpoor, Bioinspired rational design of bi-material 3d printed soft-hard interfaces, Nature Communications 14~(1) (2023) 7919.

\bibitem{conde2024multi}
F.~M. Conde, P.~G. Coelho, J.~M. Guedes, Multi-scale topology optimization of structures with multi-material microstructures using stiffness and mass design criteria, Advances in Engineering Software 187 (2024) 103566.

\bibitem{guo2025multi}
Y.~Guo, T.~Gang, S.~Cheng, Y.~Wang, L.~Chen, Multi-scale multi-material concurrent topology optimization of graded cellular structures, Engineering with Computers 41~(4) (2025) 2737--2764.

\bibitem{shimoda2024concurrent}
M.~Shimoda, J.~Fujita, M.~Al~Ali, A.~Kamiya, Concurrent multiscale and multi-material optimization method for natural vibration design of porous structures, International Journal for Numerical Methods in Engineering 125~(7) (2024) e7424.

\bibitem{wu2021topology}
J.~Wu, O.~Sigmund, J.~P. Groen, Topology optimization of multi-scale structures: a review, Structural and Multidisciplinary Optimization 63~(3) (2021) 1455--1480.

\bibitem{ma2024asymptotic}
P.-S. Ma, X.-C. Liu, X.-L. Luo, S.~Li, L.-W. Zhang, Asymptotic homogenization of phase-field fracture model: An efficient multiscale finite element framework for anisotropic fracture, International Journal for Numerical Methods in Engineering 125~(13) (2024) e7489.

\bibitem{ituarte2019DesignAdditiveManufacture}
I.~F. Ituarte, N.~Boddeti, V.~Hassani, M.~L. Dunn, D.~W. Rosen, Design and additive manufacture of functionally graded structures based on digital materials, Additive Manufacturing 30 (2019) 100839.
\newblock \href {https://doi.org/10.1016/j.addma.2019.100839} {\path{doi:10.1016/j.addma.2019.100839}}.

\bibitem{vijayakumaran2025ConsistentMachineLearninga}
H.~Vijayakumaran, J.~B. Russ, G.~H. Paulino, M.~A. Bessa, Consistent machine learning for topology optimization with microstructure-dependent neural network material models, Journal of the Mechanics and Physics of Solids 196 (2025) 106015.
\newblock \href {https://doi.org/10.1016/j.jmps.2024.106015} {\path{doi:10.1016/j.jmps.2024.106015}}.

\bibitem{wei2026MachineLearningbasedMultiscale}
D.~Wei, F.~Zeng, J.~Cui, Y.~Wang, Machine learning-based multiscale topology optimization framework for nonlinear materials, International Journal of Mechanical Sciences 314 (2026) 111380.
\newblock \href {https://doi.org/10.1016/j.ijmecsci.2026.111380} {\path{doi:10.1016/j.ijmecsci.2026.111380}}.

\bibitem{stollberg2025multiscale}
J.~Stollberg, T.~Gangwar, O.~Weeger, D.~Schillinger, Multiscale topology optimization of functionally graded lattice structures based on physics-augmented neural network material models, Computer Methods in Applied Mechanics and Engineering 438 (2025) 117808.

\bibitem{jadoon2026multiscale}
A.~A. Jadoon, A.~Tyagi, L.~R. Spencer, R.~E. Jones, M.~K. Rausch, R.~Alberdi, D.~T. Seidl, J.~N. Fuhg, Multiscale topology optimization of compressible and nearly incompressible anisotropic hyperelastic structures using physics-augmented neural networks, Computer Methods in Applied Mechanics and Engineering 461 (2026) 119270.

\bibitem{tan2026towards}
J.~Tan, G.~A. Padmanabha, S.~J. Yang, N.~Bouklas, Towards rapid constitutive model discovery from multi-modal data: Physics augmented finite element model updating (pafemu), arXiv preprint arXiv:2604.07746 (2026).

\bibitem{amos2017input}
B.~Amos, L.~Xu, J.~Z. Kolter, Input convex neural networks, in: International conference on machine learning, PMLR, 2017, pp. 146--155.

\bibitem{louizos2017learning}
C.~Louizos, M.~Welling, D.~P. Kingma, Learning sparse neural networks through {$L_0$} regularization, arXiv preprint arXiv:1712.01312 (2017).

\bibitem{kingma2014adam}
D.~P. Kingma, J.~Ba, Adam: A method for stochastic optimization, arXiv preprint arXiv:1412.6980 (2014).

\bibitem{fernandez2018semi}
F.~Fernandez, D.~A. Tortorelli, Semi-analytical sensitivity analysis for nonlinear transient problems, Structural and Multidisciplinary Optimization 58~(6) (2018) 2387--2410.

\bibitem{barzilai1988TwoPointStepSize}
J.~Barzilai, J.~M. Borwein, Two-{{Point Step Size Gradient Methods}}, IMA Journal of Numerical Analysis 8~(1) (1988) 141--148.
\newblock \href {https://doi.org/10.1093/imanum/8.1.141} {\path{doi:10.1093/imanum/8.1.141}}.

\bibitem{holmberg2013StressConstrainedTopology}
E.~Holmberg, B.~Torstenfelt, A.~Klarbring, Stress constrained topology optimization, Structural and Multidisciplinary Optimization 48~(1) (2013) 33--47.
\newblock \href {https://doi.org/10.1007/s00158-012-0880-7} {\path{doi:10.1007/s00158-012-0880-7}}.

\bibitem{svanberg1987MethodMovingAsymptotes}
K.~Svanberg, The method of moving asymptotes---a new method for structural optimization, International Journal for Numerical Methods in Engineering 24~(2) (1987) 359--373.
\newblock \href {https://doi.org/10.1002/nme.1620240207} {\path{doi:10.1002/nme.1620240207}}.

\bibitem{svanberg2002class}
K.~Svanberg, A class of globally convergent optimization methods based on conservative convex separable approximations, SIAM journal on optimization 12~(2) (2002) 555--573.

\bibitem{NLopt}
S.~G. Johnson, The {NLopt} nonlinear-optimization package, \url{https://github.com/stevengj/nlopt} (2007).

\bibitem{bendsoe1989optimal}
M.~P. Bends{\o}e, Optimal shape design as a material distribution problem, Structural optimization 1~(4) (1989) 193--202.

\bibitem{sigmund200199}
O.~Sigmund, A 99 line topology optimization code written in matlab, Structural and multidisciplinary optimization 21~(2) (2001) 120--127.

\bibitem{wang2014InterpolationSchemeFictitious}
F.~Wang, B.~S. Lazarov, O.~Sigmund, J.~S. Jensen, Interpolation scheme for fictitious domain techniques and topology optimization of finite strain elastic problems, Computer Methods in Applied Mechanics and Engineering 276 (2014) 453--472.
\newblock \href {https://doi.org/10.1016/j.cma.2014.03.021} {\path{doi:10.1016/j.cma.2014.03.021}}.

\bibitem{deng2025DatadrivenMultiscaleNonlinear}
S.~Deng, H.~D. Espinosa, W.~Chen, Data-driven multiscale nonlinear topology optimization of functionally graded soft composites under large deformations, Computational Mechanics (Dec. 2025).
\newblock \href {https://doi.org/10.1007/s00466-025-02730-1} {\path{doi:10.1007/s00466-025-02730-1}}.

\bibitem{lazarov2011FiltersTopologyOptimization}
B.~S. Lazarov, O.~Sigmund, Filters in topology optimization based on {{Helmholtz-type}} differential equations, International Journal for Numerical Methods in Engineering 86~(6) (2011) 765--781.
\newblock \href {https://doi.org/10.1002/nme.3072} {\path{doi:10.1002/nme.3072}}.

\bibitem{bredenbeck2026aerial}
A.~Bredenbeck, A.~Jadoenathmisier, S.~Hamaza, Aerial tactile perching via an anthropomorphic hand with embodied soft tactile receptors, npj Robotics 4~(1) (2026) 35.

\bibitem{mankame2002contact}
N.~D. Mankame, G.~Ananthasuresh, Contact aided compliant mechanisms: concept and preliminaries, in: International design engineering technical conferences and computers and information in engineering conference, Vol. 36533, American Society of Mechanical Engineers, 2002, pp. 109--121.

\bibitem{faltus2024ThirdMediumFinite}
O.~Faltus, M.~Hor{\'a}k, M.~Do{\v s}k{\'a}{\v r}, O.~Roko{\v s}, Third medium finite element contact formulation for pneumatically actuated systems, Computer Methods in Applied Mechanics and Engineering 431 (2024) 117262.
\newblock \href {https://doi.org/10.1016/j.cma.2024.117262} {\path{doi:10.1016/j.cma.2024.117262}}.

\bibitem{wriggers2025ThirdMediumApproach}
P.~Wriggers, J.~Korelc, {\relax Ph}.~Junker, A third medium approach for contact using first and second order finite elements, Computer Methods in Applied Mechanics and Engineering 436 (2025) 117740.
\newblock \href {https://doi.org/10.1016/j.cma.2025.117740} {\path{doi:10.1016/j.cma.2025.117740}}.

\bibitem{arruda1993three}
E.~M. Arruda, M.~C. Boyce, A three-dimensional constitutive model for the large stretch behavior of rubber elastic materials, Journal of the Mechanics and Physics of Solids 41~(2) (1993) 389--412.

\bibitem{mousavi2025chain}
S.~M. Mousavi, J.~Mulderrig, B.~Talamini, N.~Bouklas, A chain stretch-based gradient-enhanced model for damage and fracture in elastomers, Computer Methods in Applied Mechanics and Engineering 444 (2025) 118103.

\bibitem{da2021local}
G.~A. Da~Silva, N.~Aage, A.~T. Beck, O.~Sigmund, Local versus global stress constraint strategies in topology optimization: a comparative study, International Journal for Numerical Methods in Engineering 122~(21) (2021) 6003--6036.

\bibitem{senhora2020topology}
F.~V. Senhora, O.~Giraldo-Londono, I.~F. Menezes, G.~H. Paulino, Topology optimization with local stress constraints: a stress aggregation-free approach, Structural and Multidisciplinary Optimization 62~(4) (2020) 1639--1668.

\bibitem{zorzetto2020properties}
L.~Zorzetto, L.~Andena, F.~Briatico-Vangosa, L.~De~Noni, J.-M. Thomassin, C.~J{\'e}r{\^o}me, Q.~Grossman, A.~Mertens, R.~Weinkamer, M.~Rink, et~al., Properties and role of interfaces in multimaterial 3d printed composites, Scientific reports 10~(1) (2020) 22285.

\bibitem{bluhm2021InternalContactModeling}
G.~L. Bluhm, O.~Sigmund, K.~Poulios, Internal contact modeling for finite strain topology optimization, Computational Mechanics 67~(4) (2021) 1099--1114.
\newblock \href {https://doi.org/10.1007/s00466-021-01974-x} {\path{doi:10.1007/s00466-021-01974-x}}.

\end{thebibliography}
\end{document}